\documentclass[10pt]{iopart}
    
    \usepackage{rotating}
    \usepackage{tabularx}
    \usepackage{graphicx}

     \newcolumntype{Z}{>{\centering\arraybackslash}X}
        
\begin{document}

    \title[Measurement of ion and electron drift velocity and electronic attachment for air-IC]{Measurement of ion and electron drift velocity and electronic attachment in air for ionization chambers.}{} 
    
    \author{G. Boissonnat, JM. Fontbonne, J. Colin, A. Remadi, S. Salvador}
    	\address{LPC Caen (Normandie Univ-ENSICAEN-UNICAEN-CNRS/IN2P3) 6 Boulevard Mar\'echal Juin, 14050 Caen, France}
    \ead{fontbonne@lpccaen.in2p3.fr}
    
    \vspace{10pt}
    \date{\today}
    
    \begin{abstract}
    
    Air-ionization chambers have been used in radiotherapy and particle therapy for decades. However, fundamental parameters in action in the detector responses are sparsely studied. In this work we aimed to measure the electronic attachment, electrons and ions mobilities of an ionization chamber (IC) in air. The main idea is to extract these from the actual response of the IC to a single ionizing particle in order to insure that they were measured in the same condition they are to be used while neglecting undesired phenomena: recombination and space charge effect. The non-standard signal shape analysis performed here were also confronted to a more standard drift chamber measurements using time-of-flight. It was found that both detectors displayed compatible results concerning positive and negative ions drift velocities where literature data is well spread out. In the same time, electron attachment measurements sit in the middle of known measurements while electron drift velocities seemed to show an offset compared to the well known and modeled data.
    
    \end{abstract}
    
    \section{Introduction}
    \label{sec_intro}
    
    For years, air ionization chambers have been the gold standard in beam monitoring and dosimetry for radiotherapy and particle therapy. They have the advantages to be relatively simple, very resistant to radiations, to have little impact on the beam, and to have linear dose responses in a wide range of intensity and irradiation modalities. In addition, the use of air as ionizing medium is very convenient in the medical context while it limits the number of apparatus needed and safety risks.  The increasing beam intensity observed in latest development like IBA's SuperConducting SynchroCyclotron (S2C2)~\cite{IBA} for proton therapy, put the air ionization chambers at the very end of their linear operating range. In fact, the tremendous amount of ionization pairs produced in the gas causes space charge effects, amplifying charge recombination and therefore decreases the detector efficiency. To allow for correct dose measurement at high intensity, one can use Boag's calculations on detector efficiency~\cite{Boag} to correct from charge loss. Unfortunately, those theories tend to overly simplify the problem by using effective parameters which need to be adjusted on data and cannot be expressed as a function of gas properties (electron and ion drift velocities, recombination and attachment coefficients), chamber design (electric field and gap size) and beams characteristics (spatial and temporal beam shape). Therefore, there is not any simple way to adapt an ionization chamber design to a specific application beforehand. 
    
    This study is motivated by previous developments done at LPC Caen on ionization chambers for particle therapy applications \cite{IC23} and by the need to accurately simulate the efficiency of a given ionization chamber to a known irradiation condition. While every phenomenon occurring in an air-ionization chamber seems to be known, the lack of agreement on air swarm parameters in the literature renders any modeling attempt unlikely. The purpose here is therefore to measure electron and ion drift velocities along with electronic attachment time in air, in conditions adapted to medical ionization chambers, to be used in later simulations. 
    
    The drift velocity $\vec{v}$ is usually related to the electric field $\vec{E}$ through the mobility $K$ (Eq.~\ref{eqVelocity}) using notations from \cite{Ellis76}. It implies that $K$ is positive for cations and negative for both anions and electrons. To enable comparisons with measurements done in several density conditions, the standard mobility $K_0$ is used and defined as the mobility at $273.15\,^{\circ}{\rm K}$ and 1013.25~hPa as in Eq.~\ref{StandardMobility} where $N$ is the gas density in number of neutral particle per unit of volume. It is also common to use the reduced electric field ($E / N$). However, one should not assume that $K_0$ is independent from the electric field or from the condition at which it was measured. 
    
    \begin{equation}
    \vec{v} = K \cdot \vec{E} 
    \label{eqVelocity}
    \end{equation}
    
    \begin{equation}
    K_0 = \frac{N K}{N_0} = K \left( \frac{273.15}{T} \right) \left( \frac{P}{1013.25} \right)  
    \label{StandardMobility}
    \end{equation}
    
    Moreover, the electronic attachment phenomenon is the reaction of electrons with neutral molecules or atoms from the gas leading to the formation of anions. As described in Eq.~\ref{NbodyAttachment}, they can be formed either through two-body or three-body attachment processes \cite{Kinetic}.

    \begin{eqnarray}
    e^{-}+ O_2 \rightarrow O_2^{-} \nonumber \\
    e^{-}+ O_2 + O_2 \rightarrow O_2^{-}  + O_2 \nonumber\\
    e^{-}+ O_2 + N_2\rightarrow O_2^{-}  + N_2 
    \label{NbodyAttachment}
    \end{eqnarray}
    
    In air, electronic attachment can be expressed as the Townsend attachment coefficient ($\eta$ in m$^{-1}$, see \cite{Pancheshnyi}) as presented in Eq.~\ref{TownsendAttachentEq} where $k_2$, $k_{3_{O_2}}$, $k_{3_{N_2}}$ are the kinetic constant rates of attachment for two-body attachment and three-body attachment using either $O_2$ or $N_2$ as third body of the reaction, while $v_{e^{-}}$ is the electron drift velocity. Here, we define the attachment time $T_a$ as the negative ions formation rate (Eq.~\ref{AttachementTime}), corresponding to the attachment parameter directly measured in this study.
    
    \begin{equation}
    \eta = \frac{ k_2 \left[ O_2 \right] + k_{3_{O_2}} \left[ O_2 \right] \left[ O_2 \right] + k_{3_{N_2}} \left[ O_2 \right] \left[ N_2 \right] } {v_{e^{-}}}
    \label{TownsendAttachentEq}
    \end{equation}
    
    \begin{equation}
    T_a = \frac{1}{ \eta \cdot v_{e^{-}} }
    \label{AttachementTime}
    \end{equation}

    \section{Material and methods}
    
    In order to obtain measurements of drift velocities and electronic attachment times relevant to medical ionization chambers, it was decided to use gaseous detectors operating in similar conditions while preventing unwanted phenomena such as recombination and space charge effects to occur. In that framework, two dedicated setups were developed in-house. First, a parallel plate chamber named MICMAC and second, a drift chamber (used in ionization chamber mode) named MAGIC. Both of them followed the same principle: 5.5~MeV $\alpha$ particles (emitted by a 44~kBq $^{241}{\rm Am}$ source electro-deposited on a metal plat) crossed the detector active area, ionizing the gas and ending its path in a scintillator whose scintillation was detected by a PhotoMultiplier Tube (PMT) and therefore triggering the signal acquisition. The detection of individual $\alpha$ particles allowed us to create a uniformly ionized track containing only 15.000 electron-ions pairs (2.4~fC) and to operate both detectors in pulse-mode. 
    
    In this study, the electric field in the active area of both detectors were supposed homogeneous and undisturbed by the charges in motion. For both detectors, measurements were done in the laboratory conditions (291$^{\circ}$K and 1015~hPa) for both air and pure nitrogen gas. While our purpose was to obtain measurements adapted to air ionization chambers, measurements in nitrogen were used to validate our signal shape analysis methods as the corresponding literature data is much better furnished in nitrogen than in air.
    
    In addition, the difference in drift velocities between electrons and ions (about three order of magnitude) forced us to use different electronic acquisition setups to optimize measurements for both electronic and ionic signal components. Nevertheless, even after amplification, measured signals were to small in amplitude compared to the preamplifier noise to be usable individually. For both setups, and each condition, the induced signal was averaged over few thousands detected $\alpha$ particles.

    \subsection{The ionization chamber MICMAC}
    
\begin{figure}[!ht]
    \begin{center}
    \includegraphics[angle=0, width=0.95\columnwidth]{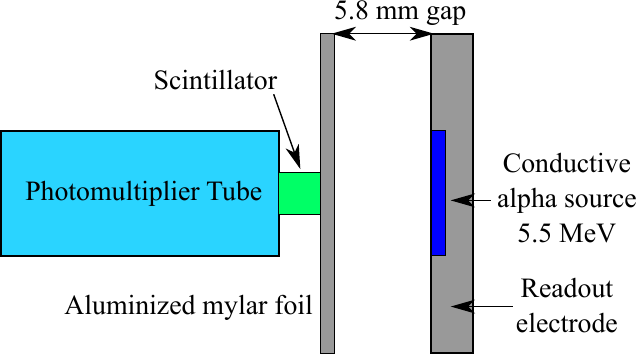}
    \end{center}
    \caption{Schematic of the MICMAC ionization chamber.}
    \label{MicMacSchem}
    \end{figure}

    The detector MICMAC was a 5.8$\pm$0.03~mm-gap parallel plate ionization chamber (Fig.~\ref{MicMacSchem}). As with every other ionization chambers, ionized charges were put into motion through the electrical potential difference between electrodes. The specificity of this setup was that the readout electrode was the conductive $\alpha$ source itself while the high voltage electrode was constituted of a 2.5~$\mu$m thick aluminized mylar foil enabling the $\alpha$ particle to get through and be detected by the scintillation detector placed behind it (as presented in  Fig.~\ref{MicMacSchem}). 
    
    When an $\alpha$ particle went through the air gap, two groups of charges (electrons and cations) moving along the same line in opposite directions are created. The induced signal was modeled as a function of their speed and attachment time using the Ramo-Shockley theorem \cite{Ramo}. The induced current is presented in Eq.~\ref{EqRamo}, where each individual ionized particle $i$, of charge $q_i$ and speed $\vec{v_i}$ moved in the applied electric field  $\vec{E}$ with $\vec{E^{*}}$ the corresponding virtual electric field. This virtual electric field was defined as the electric field that would exist if the measurement electrode was put at 1~V while all other electrodes were placed at the ground potential.
    
    \begin{equation}
    I(t)=\Sigma_i q_i \cdot \vec{v_i} (\vec{E},t) \cdot \frac{\vec{E^{*}}}{\textrm{1~V}}
    \label{EqRamo}
    \end{equation}
    
    In the case of MICMAC, the virtual electric field $\left\| \vec{E^{*}} \right\|$ was equal to $1~\textrm{V}/d$ where $d$ was the electrode spacing while its direction was collinear with the direction of the moving charges. In addition, one should note that both electron and cation-induced signals had the same polarity as species had opposite signs while moving in opposite directions. Therefore, when using the MICMAC setup, species were only differentiated by their drift time. 
    
    
    \subsubsection{Experimental conditions in air and nitrogen.}
    
    Measurements were performed using applied voltage from 100 to 4000~V at 291$\pm$2.9$^{\circ}$K and 1015$\pm$8.6~hPa. As the PMT was relatively close to the measurement electrode it produced electronic noise when triggering. To remove this noise, every measurement presented here corresponds to the difference between data obtained at the same applied voltage but in opposite polarities. The in-house build digital acquisition system FASTER \cite{refFaster} was used. Its oscilloscope mode registered a full window of 700 samples around the triggering time. Depending on the species measured, either ions or electrons, an adapted sampling time ($T_s$) was used: 2~ns for electrons and 512~ns averaging for ions. 
    
    \subsubsection{Signal shape modeling in nitrogen filled MICMAC.}
    
As attachment does not occur in nitrogen, no anion can be present and the measured current can be divided in two: the electron-induced current $i_{N_2}^{~e^-}$ and the cation-induced current $i_{N_2}^{~+}$. With a uniform electric field across the detector, all charges of the same species are moving at the same drift velocity $v_d$ . This drift velocity is equal to the drift distance (inter-electrode spacing) divided by the drift time $t_d$ ($t_{d_{e^-}}$ for electrons and $t_{d_+}$ for cations, respectively). Using this notation to replace the velocity in the Ramo-Shockley theorem (Eq.~\ref{EqRamo}), we obtain Eq.~\ref{NitrogenEqMicMacThEle} and Eq.~\ref{NitrogenEqMicMacThCat} where the only parameters involved are the drift times and the number of electron-cation pairs created by the ionization $Q_0$.

    \begin{eqnarray}
    i_{N_2}^{~e^-} (t) & =   \frac{ Q_0}{t_{d_{e^-}}} \left(1-\frac{t}{t_{d_{e^-}}}\right) \label{NitrogenEqMicMacThEle} \\
    i_{N_2}^{~+} (t) & =  \frac{Q_0}{t_{d_+}} \left(1-\frac{t}{t_{d_+}}\right) \label{NitrogenEqMicMacThCat} 
    \end{eqnarray}
    
    In practice, measured signals (presented in Eq.~\ref{NitrogenEqMicMacThSamplingFast} and \ref{NitrogenEqMicMacThSamplingSlow}) contain only the part of the information accessible at the used sampling time. With the fast sampling rate ($T_s = 2$~ns), the amplitude of $i_{N_2}^{~+}$ is too small to be measured, therefore only the electronic component $i_{N_2}^{~e^-}$ is presented on the measured signal $S_{N_2}^{T_s = 2~\textrm{ns}} $. Similarly, at the slower sampling rate ($T_s = 512$~ns), the measured signal $S_{N_2}^{T_s = 512~\textrm{ns}}$ contains the electron-induced current as a $\delta$-Dirac function of amplitude $Q_{N_2}^{e^-}$ which is equal to the integral of $i_{N_2}^{~e^-}(t)$ (as expressed in Eq.~\ref{NitIntEle}).  
    
    \begin{eqnarray}
    S_{N_2}^{T_s = 2~\textrm{ns}} (t) & = i_{N_2}^{~e^-} (t) \label{NitrogenEqMicMacThSamplingFast} \\
    S_{N_2}^{T_s = 512~\textrm{ns}} (t) & = Q_{N_2}^{e^-} \cdot \delta(t) +i_{N_2}^{~+} (t)  \label{NitrogenEqMicMacThSamplingSlow} \\
    Q_{N_2}^{e^-} & =\int_{0}^{t_{d_{e^-}}} i_{N_2}^{~e^-} (t) \cdot \mathrm{d}t = \frac{Q_0}{2}  \label{NitIntEle} 
    \end{eqnarray}
    
    As a result, in nitrogen, electron mobilities can be extracted from $S_{N_2}^{T_s = 2~\textrm{ns}} $ through $t_{d_{e^-}}$ and cation mobilities from $S_{N_2}^{T_s = 512~\textrm{ns}}$ using the corresponding drift time ($t_{d_+}$).
    
    \subsubsection{Signal shape modeling in air-filled MICMAC.}\label{TaOverVeRatio}
    
    In air, electronic attachment phenomenon transforms electrons in anions as they are moving to their collection electrode, therefore decreasing the electron-induced signal and creating an anion-induced signal. The electron-induced component $i_{Air}^{~e^-}$ presented in Eq.~\ref{AirEqMicMacThEle} decreases exponentially (with a time constant equal to the attachment time $T_a$) due to attachment compared to the signal seen in nitrogen (Eq.~\ref{NitrogenEqMicMacThEle}). If the cation signal $ i_{Air}^{~+}$ itself is not affected by the attachment (Eq.~\ref{AirEqMicMacThCat}), the signal resulting from the formation of anions $i_{Air}^{~-}$ (Eq.~\ref{AirEqMicMacThAn}) is much less straightforward to model. In fact, it depends on the number of electron undergoing attachment as a function of time and position along the drift line. Due to their difference in velocity, we can assume that the produced anions are stationary while electrons are being collected ($t<t_{d_{e^-}}$). Thus, the number of anion ($\rho_{Air}^{~-}$) produced at a given position $x$ is related to the time needed by electrons located between that position and the anode (at $x$=0) to move past this point (see in Eq.~\ref{eqDistAnionMICMAC}).
    
    \begin{eqnarray}
    i_{Air}^{~e^-} (t)& = \frac{Q_0}{t_{d_{e^-}}}  \left(1-\frac{t}{t_{d_{e^-}}}\right) \cdot exp \left(-\frac{t}{T_a}\right)  \label{AirEqMicMacThEle} \\
    i_{Air}^{~+} (t) & =  \frac{Q_0}{t_{d_+}}  \left(1-\frac{t}{t_{d_+}}\right) \label{AirEqMicMacThCat} \\
    i_{Air}^{~-} (t)& = \frac{ Q_0}{t_{d_-}}  \left[1-\frac{t}{t_{d_-}} \right. \label{AirEqMicMacThAn} \\
    &- \frac{T_a}{t_{d_{e^-}}} \left. \left\{ 1- exp\left( - \frac{t_{d_{e^-}}}{T_a} \left( 1-\frac{t}{t_{d_-}} \right)  \right) \right\} \right] \nonumber
    \end{eqnarray}
    
    \begin{eqnarray}
    \rho_{Air}^{~-} (x,~t=t_{d_{e^-}})& = \rho_{Air}^{~e^-} (t=0) \label{eqDistAnionMICMAC} \\
    & \times \left\{ 1- exp \left(-\frac{ (d-x) \cdot t_{d_{e^-}} }{d \cdot T_a}\right) \right\} \nonumber
    \end{eqnarray}
    
    Similarly to measurements in nitrogen, the ones performed in air at high sampling rates signal $S_{Air}^{T_s = 2~\textrm{ns}}$ contain only electron-induced current $i_{Air}^{~e^-}$ (Eq.~\ref{AirEqMicMacThSamplingFast}), while slower sampling rate signals $S_{Air}^{T_s = 512~\textrm{ns}}$ are composed of both cation and anion-induced current (Eq.~\ref{AirEqMicMacThSamplingSlow}) along with a $\delta$-Dirac function containing the integral of the electronic signal $Q_{Air}^{e^-}$ (Eq.~\ref{AirIntEle}). 
    
    \begin{eqnarray}
    S_{Air}^{T_s = 2~\textrm{ns}} (t) & = i_{Air}^{~e^-} (t) \label{AirEqMicMacThSamplingFast} \\
    \nonumber \\
    S_{Air}^{T_s = 512~\textrm{ns}} (t) & = Q_{Air}^{e^-} \cdot \delta(t)+ i_{Air}^{~-} (t) + i_{Air}^{~+} (t) \label{AirEqMicMacThSamplingSlow} 
    \end{eqnarray}
    
    \begin{eqnarray}
    Q_{Air}^{e^-} = Q_0 \cdot \frac{T_a}{t_{d_{e^-}}}  \left[1-\frac{T_a}{t_{d_{e^-}}}  \left\{ 1-exp \left(-\frac{t_{d_{e^-}}}{T_a} \right) \right\} \right]  \label{AirIntEle} 
    \end{eqnarray}
    
    As a result, in air, electron mobilities can be extracted from $S_{Air}^{T_s = 2~\textrm{ns}} $ through $t_{d_{e^-}}$, while both anion and cation mobilities are evaluated from $S_{Air}^{T_s = 512~\textrm{ns}}$ using their respective drift times $t_{d_+}$ and $t_{d_-}$. In addition, the attachment coefficient can be obtained from measurements at both sampling times, as the attachment time $T_a$ at the faster sampling rate, but only as the ratio $T_a$ over $t_{d_{e^-}}$ at the slower one (ratio used in both $i_{Air}^{~-} (t)$ and $Q_{Air}^{e^-}$).

    \subsubsection{Oscilloscope-induced bias correction.}
    
\begin{figure}[!ht]
    \begin{center}
    (a)\includegraphics[angle=-90, width=0.95\columnwidth]{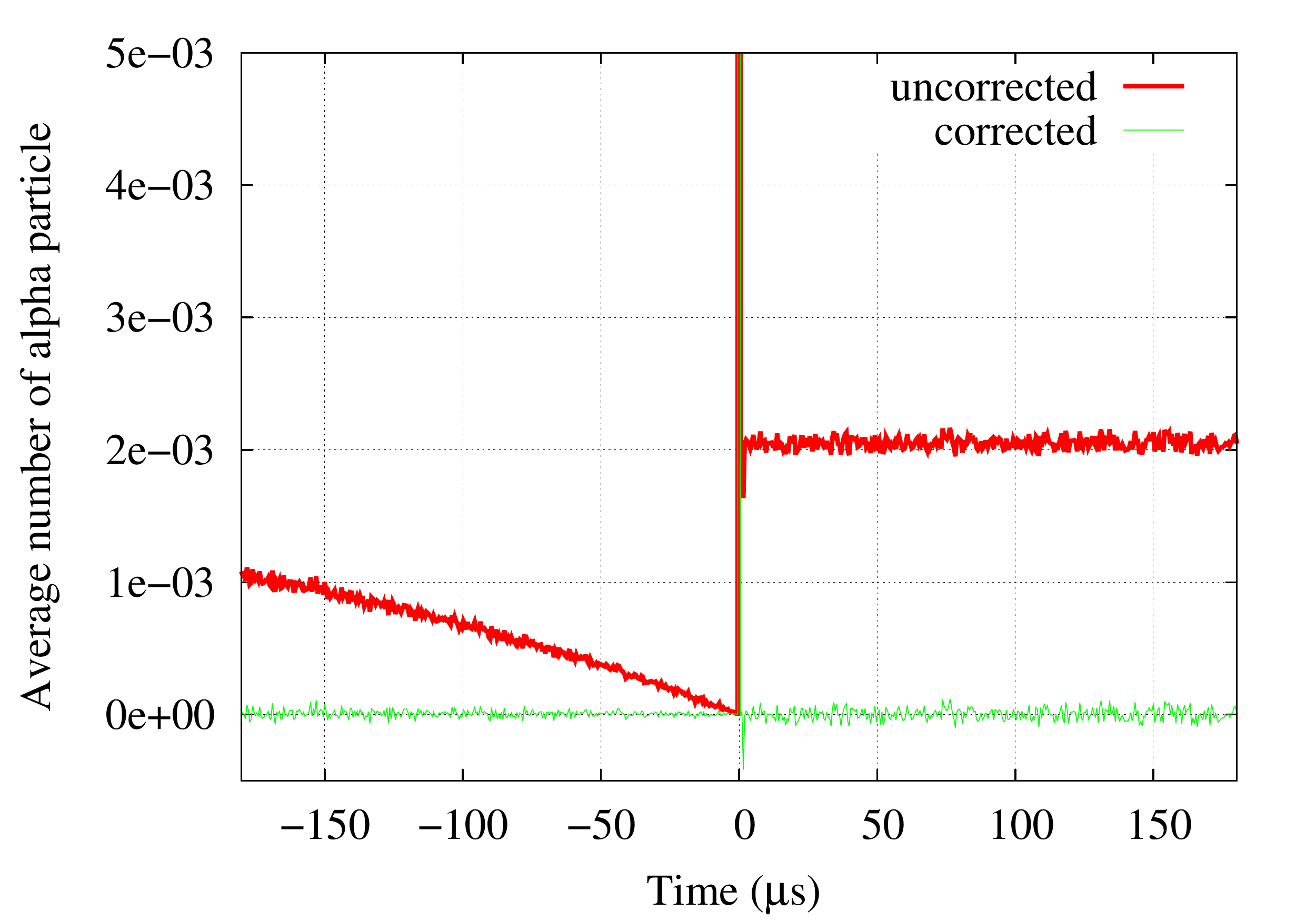}
    (b)\includegraphics[angle=-90, width=0.95\columnwidth]{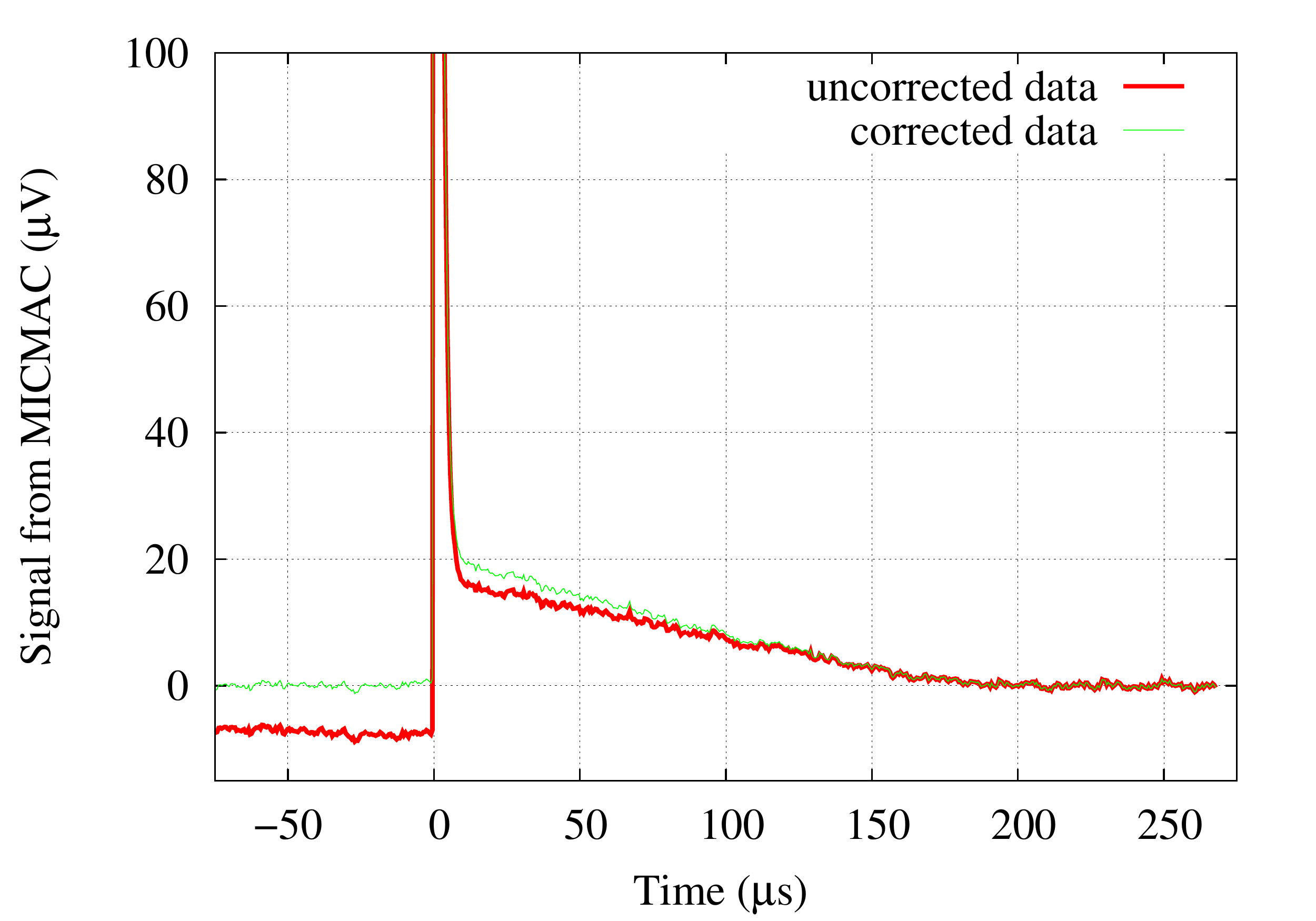}
    \end{center}
    \caption{Oscilloscope windowing effect before and after correction using the bilinear Z-Transform function. (a)~Theoretical shape of the effect (Eq.~\ref{EqProbaPM}). (b)~Measurement for ion in nitrogen (at 4~kV).}
    \label{MicMacPM2}
    \end{figure}

    In practice, a measured signal cannot be directly compared to its model counterpart, as experimental effects have to be taken into account. The first one is related to the trigger-based averaging method used for noise reduction. Indeed, while signals were averaged, the registered signal does not correspond to a single event but a time window containing signal samples before and after the triggering time. Their might have been multiple $\alpha$ particles detected by the photomultiplier tube in the same time window but only one triggered the acquisition. Each time window is followed by a dead time during which secondary non-triggering $\alpha$ particles may cross the detector. Those same non-triggering events might still be registered as part of the time window of a later triggering event. Therefore, when averaging, the measured signal corresponds to the signal induced by a single $\alpha$ particle in the ionization chamber convolved by the mean number of event in a time sample as a function of its position in that time window. This last function can be obtained by a simple simulation and is presented in Fig.~\ref{MicMacPM2}(a) along with its equation in Eq.~\ref{EqProbaPM} where $\tau$ is the average time between two events (400~$\mu$s) and $T_s$ the sampling time (512~ns).
    
    \begin{eqnarray}
    P(t) =  \left\{
                    \begin{array}{lc}
                     t < 0,&   \frac{T_s}{\tau}   \left( 1 - exp \left(  \frac{t}{\tau} \right) \right)
    \\ t = 0,&1
    \\ t > 0, & \frac{T_s}{\tau} 
      \end{array}
                  \right.
    \label{EqProbaPM}
    \end{eqnarray}
    
    In order to correct for the oscilloscope windowing effect, first, the offset was measured using the last samples in the window. Then, the exponential part was removed using numerical inverse filtering (calculated with a bilinear Z-transform) as presented in Eqs.~\ref{invFiltering} and~\ref{invFilteringCoeff} where $E$ is the measured signal and $S$ the inverse filtered one. In order to simplify the expression of the inverse filter, the time has been reversed, consequently a given sample $S[n]$ becomes a function of the following sample $S[n+1]$ in Eq.~\ref{invFiltering}.
    
    \begin{equation}
    S[n]= - \frac{( a_0\cdot E[n] + a_1\cdot E[n+1] + b_1\cdot S[n+1] )}{b_0}
    \label{invFiltering}
    \end{equation}
     
    \begin{eqnarray}
                    \begin{array}{lcr}
     a_0 = 2\tau + T_s & ~~~ & a_1 = T_s - 2\tau 	\\ 
     b_0 = 2 - 2\tau  + T_s / \tau & ~~~ & b_1 = 2\tau  -2 +  T_s / \tau
       \end{array}
    \label{invFilteringCoeff}
    \end{eqnarray}
    
    Fig.~\ref{MicMacPM2} shows the results of this correction method applied on the calculated average number of $\alpha$ particle (a) as a function of time (when trigged at t=0~s) and on measurements (b). 
    
    This effect can be neglected for electron measurements as $T_s / \tau$ ratio represents 5$\times$10$^{-6}$ compared to the 1.28$\times$10$^{-3}$ for ion measurements. 
    
    \subsubsection{Electronic low-pass filtering corrections.}\label{MICMAClowPassFilter}
    
    After correction of the oscilloscope-induced bias and before adjusting the model on the data, low-pass filters, inherent to the current preamplifiers used, had to be taken into account. As it was not possible to measure the full electronic response function, it was decided to make this correction part of the model adjustment process.
    
    As two different electronics were used depending on the sampling time, two different filters were considered. The low-pass filter dedicated to ion measurements ($T_s=512$~ns) was modeled as a simple second order low-pass filter. For electron signal measurements, standard low-pass filter failed to reproduce the raising shape of the signal. It was therefore replaced by a unilateral causal digital filter using the 14 previous time samples weighted without constraint.
    
    
    \subsection{The drift chamber MAGIC}
    
\begin{figure}[!ht]
    \begin{center}
    \includegraphics[angle=0, width=0.95\columnwidth]{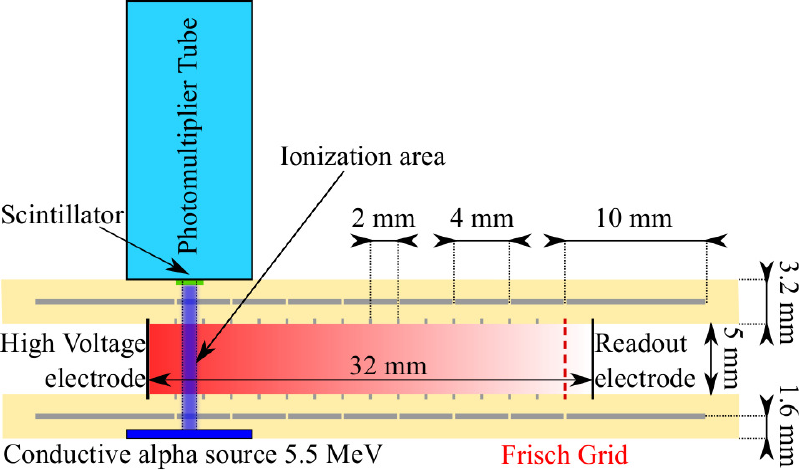}
    \end{center}
    \caption{Schematic of the MICMAC ionization chamber.}
    \label{MagicSchem}
    \end{figure}

    If the overall setup for MAGIC was very similar to the MICMAC one, its operating principale was much closer to standard drift tubes~\cite{DriftTube}. Here, the electric field was perpendicular to the ionization direction and only charges reaching the area between the collection electrode and the Frisch grid were measured (as presented in Fig.~\ref{MagicSchem}). As a result, positive and negative charges were acquired separately depending on the polarity chosen for the high voltage electrode. If MICMAC did not allow to fully de-correlate cations from anions, MAGIC enabled it by design. The drift velocity being measured through time-of-flight, the uncertainty on the ionization point of origin was reduced by limiting the area of ionization using 1~mm diameter collimators on both the source and the scintillator. Overall, the distance between the high voltage electrode and the ground electrode was 32$\pm$0.3~mm while the distance of flight was 29$\pm$0.3~mm. 
    
\begin{figure}[!ht]
    \begin{center}
    \includegraphics[angle=0, width=0.95\columnwidth]{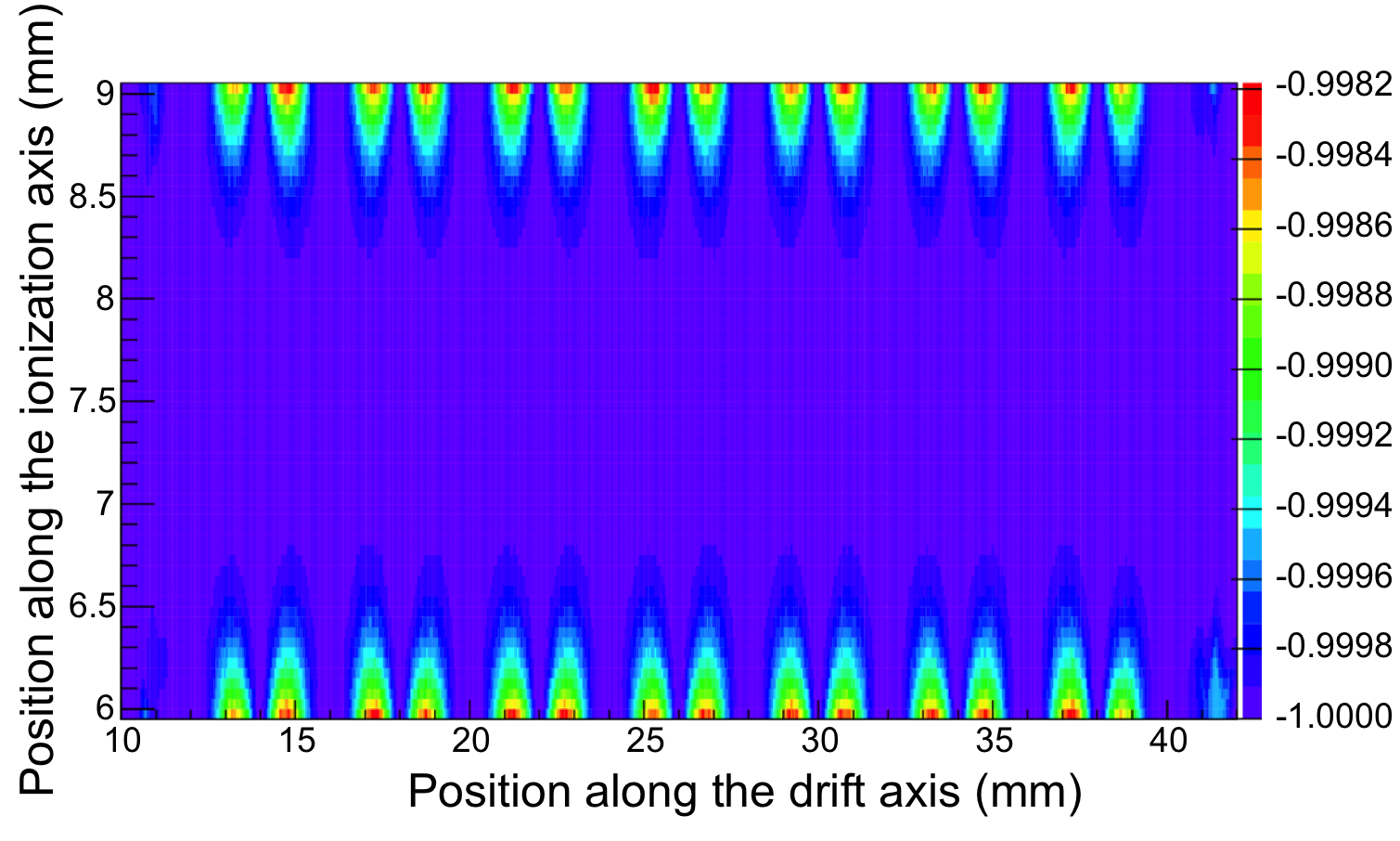}
    \end{center}
    \caption{Electric field along the drift axis normalized by the electric field norm.}
    \label{MagicField}
    \end{figure}

    In order to maintain a homogeneous electric field free from PMT-induced deformation across the drift area, MAGIC was built using two triple layer Printed Circuit Board (PCB) imbedded with an electrode pattern, one for each side of the chamber. A first layer of fifteen 2~mm spaced-copper strips was used to apply a gradient of electrical potential and a second array of nine larger strips (located in the middle layer of each PCB) was used for isolation. The overall design was validated with the finite element calculation software FreeFem++~\cite{FreeFem}, highlighting a lateral electric field distortion bellow 1\% on the 3~mm at the center of the 5~mm air gap (Fig.~\ref{MagicField}). Those 3~mm correspond to the part of the collection electrode wired to the readout electronic while the rest was directly put at ground potential.

    MAGIC was used in air and nitrogen at 291$\pm$2.9$^{\circ}$K and 1015$\pm$8.6~hPa and a high voltage ranging from 2 to 7~kV in both polarities. The acquisition itself was performed using a LeCroy WaveRunner 610Zi oscilloscope~\cite{LeCroy} with a 250~kHz sampling rate ($T_s$ = 4~$\mu$s).
    
    \subsubsection{Measurement analysis.}
     
    Compared to MICMAC measurement analysis, where signal shape analysis was mandatory to separate electron, anion and cation components, MAGIC analysis simply consists in dividing the time-of-flight by the distance between the ionization area and the collection area, the studied species being selected through the applied voltage polarity.
     In addition, the oscilloscope-induced bias was negligible here while the number of events per second was very low thanks to the collimation. 
    
    Similarly to the case of MICMAC, the signal was amplified and averaged to reduce the electronic noise. For the readout, a current preamplifier was used. Its time constant $\tau_{RC}$ was measured at 84~$\mu$s using the detector response to electron in nitrogen which is equivalent to a $\delta$-Dirac function at the considered sampling rate. 
     
    To correct the signal shape from the preamplifier transfer function, inverse filtering was applied using the Z-tranform as described in  Eq.~\ref{invFilteringMagic} and \ref{invFilteringCoeffMagic} where $E$ is the measured signal, $S$ the inverse filtered signal and $F_s$ the sampling frequency.

    \begin{eqnarray}
                    \begin{array}{l}    
    S[n]= \alpha \cdot E[n] + \beta \cdot E[n-1]  \label{invFilteringMagic} 
                    \end{array}\\
                    \begin{array}{lcr}    
     a_0 = 2\tau + T_s & ~~~ & a_1 = T_s - 2\tau 	\\  
     \alpha = 1+ \tau_{RC} \cdot F_s  & ~~~ & \beta =   \tau_{RC} \cdot F_s   
    \label{invFilteringCoeffMagic}
                    \end{array}
    \end{eqnarray}

    Because inverse filtering introduces a lot of noise, the inverse filtered signal was smoothed using a moving average method with the five previous and following samples. The drift time was finally computed as the difference between the PMT triggering time and the temporal centroid of the signal.
     
    
    \section{Results and discussion.}
    
    In this part, results are first presented as measurements and fitted curves for both setups. The actual swarm parameters values for both air and nitrogen will then be presented along with data from the literature.
    \subsection{MICMAC results}
    
    \subsubsection{Ion drift velocity.}
    
\begin{figure}[!h]
    \begin{center}
    (a)\includegraphics[angle=-90, width=0.95\columnwidth]{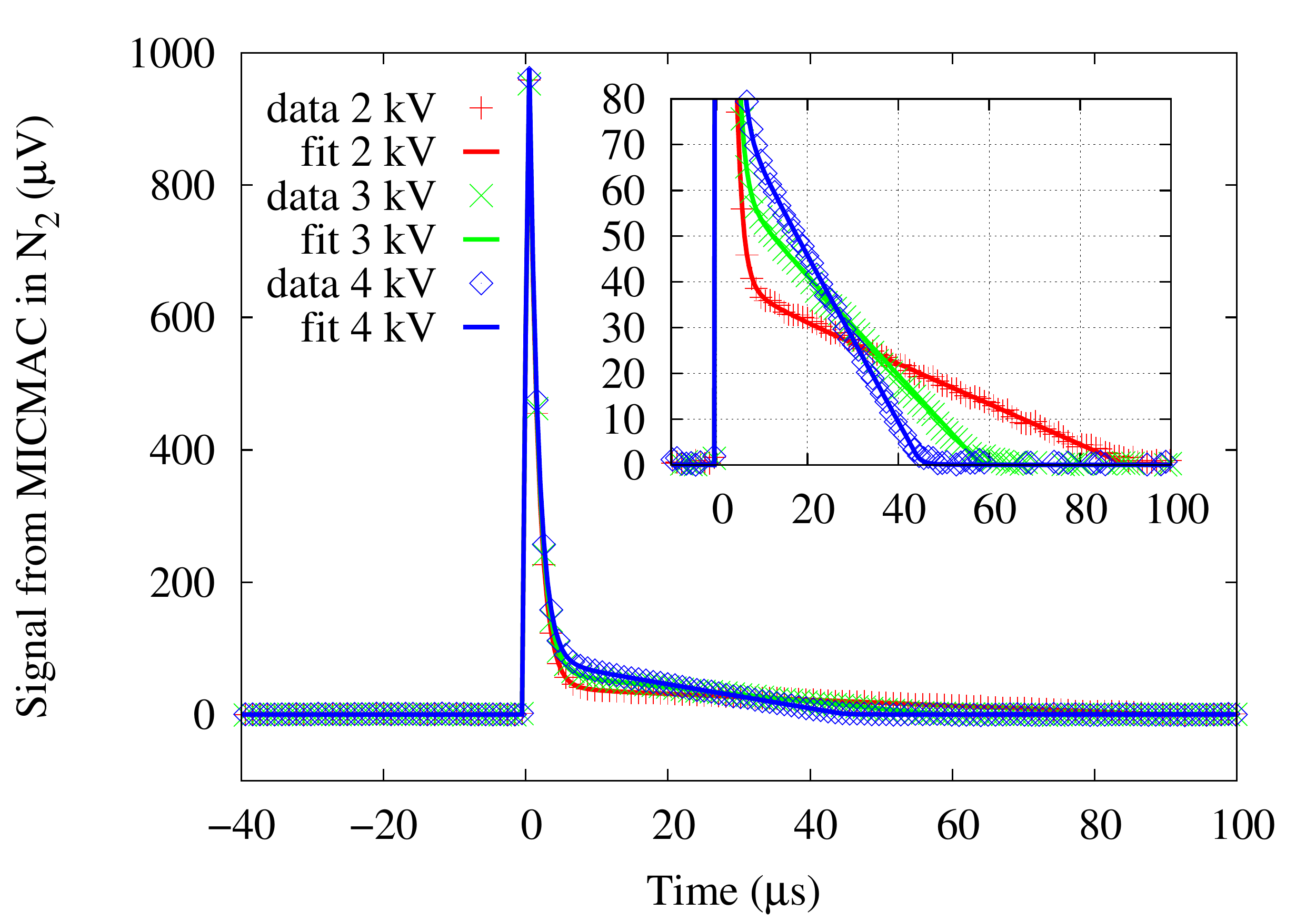}
    (b)\includegraphics[angle=-90, width=0.95\columnwidth]{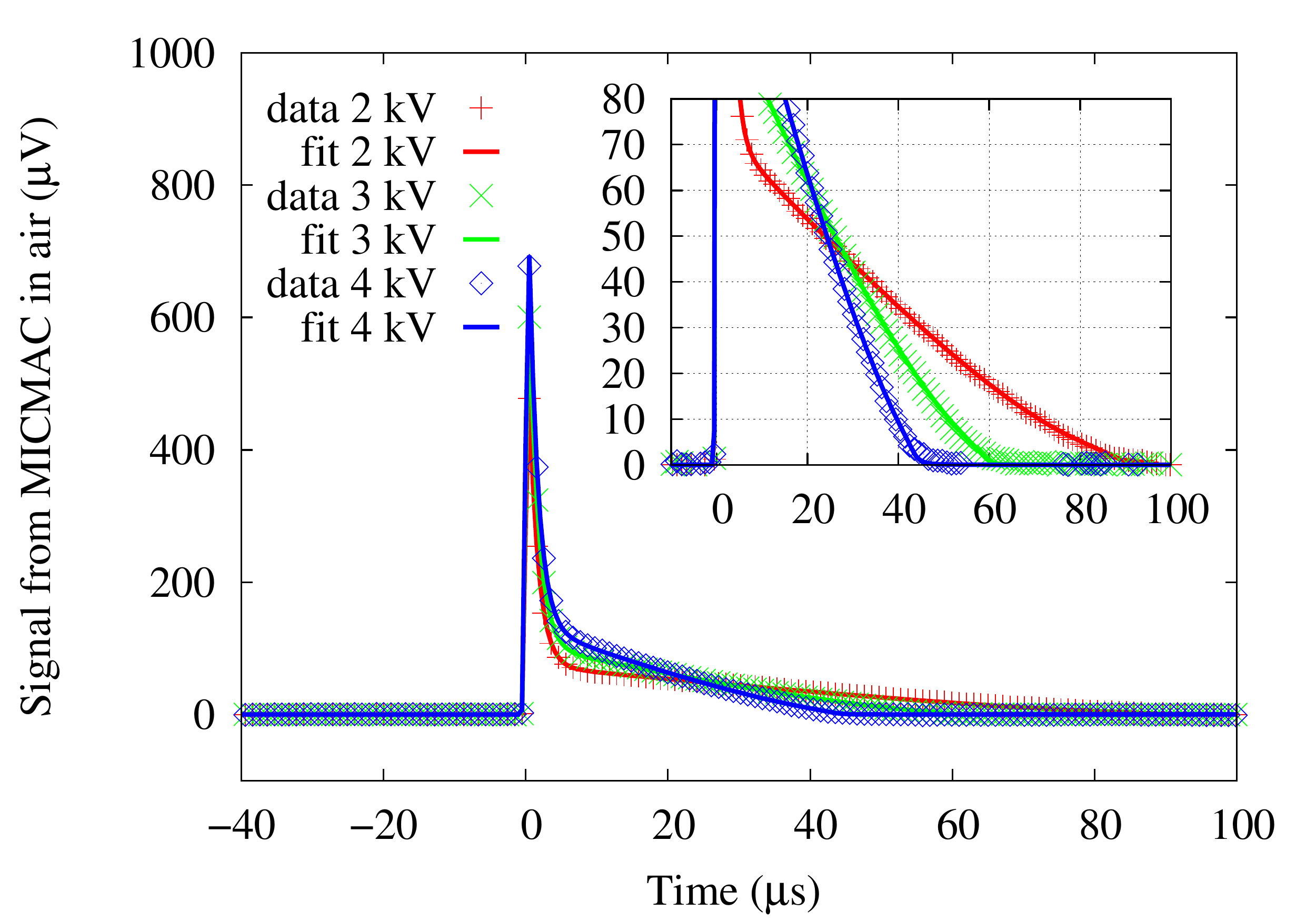}
    \end{center}
    \caption{Measurements and fitted curves for ions in (a) nitrogen and (b) air as a function of time for different applied voltages with full and zoomed signal to see both electron and ion components.}
    \label{MicMac_Ion}
    \end{figure}

    Figure~\ref{MicMac_Ion} shows MICMAC data for ion velocity measurements in nitrogen and air (corrected for the offset and oscilloscope bias) and their fitted curve using Eq.~\ref{AirEqMicMacThSamplingSlow} and Eq.~\ref{NitrogenEqMicMacThSamplingSlow} for three high voltages. The corresponding data are presented in~Table~\ref{table:MICMACIonRes} (full data presented in Annexe~\ref{table:MICMAC_Pos} and \ref{table:MICMAC_Neg}).

    Fits were performed using the same low-pass filter for every measurement (in both air and nitrogen) as well as the same number of pairs produced for each ionization regardless of the applied voltage, then individual drift and attachment times (in air only) were used for each measurement.

    \begin{table}[!ht]
    \caption{MICMAC ion drift velocity and attachment time results (ion adapted measurements)}
    \centering
    \footnotesize
    \label{table:MICMACIonRes}
    \renewcommand{\arraystretch}{1.3}
    {
    	\begin{tabularx}{\linewidth}{ZZZZZZZZZZZ}
    	\hline
    	\hline
    	HV	& $v_{d_{+}}^{N_2}$ & $v_{d_{+}}^{Air}$  & $v_{d_{-}}^{Air}$ & $T_a$ 	 \\
    	\tiny (V)	&  \tiny(mm$\cdot$s$^{-1}$) & \tiny (mm$\cdot$s$^{-1}$) & \tiny (mm$\cdot$s$^{-1}$)	 &  \tiny(ns) 	 \\
    	\hline
    	2000 	& 6.53e4 			& 6.43e4			& 7.42e4			& 70.7\\
    	3000 	& 1.02e5			& 9.79e4			& 1.11e5			& 80.5\\
    	4000 	& 1.33e5			& 1.34e5			& 1.47e5			& 84.8\\
    	\hline
    	\hline
    	\end{tabularx}
    }
    \end{table}
    
    This measurements highlight that cations drift velocities in air and nitrogen seem to be very similar (less than 5~$\%$ apart) while anion drift velocities tend to be about 10~\% higher. In addition, ions drift velocities are displaying a linear increase with the applied voltage. Finally, MICMAC attachment time measured values are slowly increasing with the electric field in the considered range.

    \subsubsection{Electron drift velocity.}
    
\begin{figure}[!h]
    \begin{center}
    (a)\includegraphics[angle=-90, width=0.95\columnwidth]{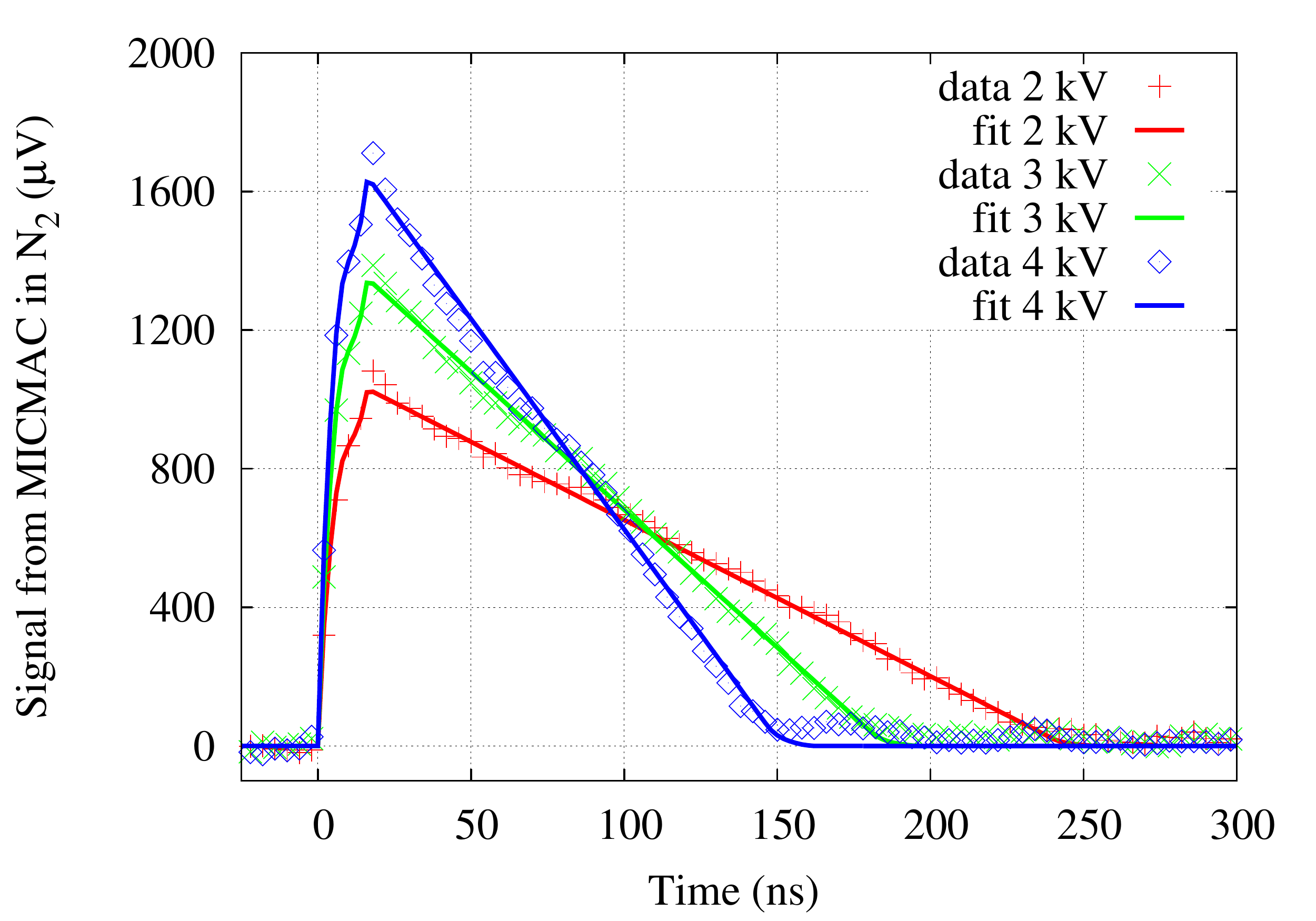}
    (b)\includegraphics[angle=-90, width=0.95\columnwidth]{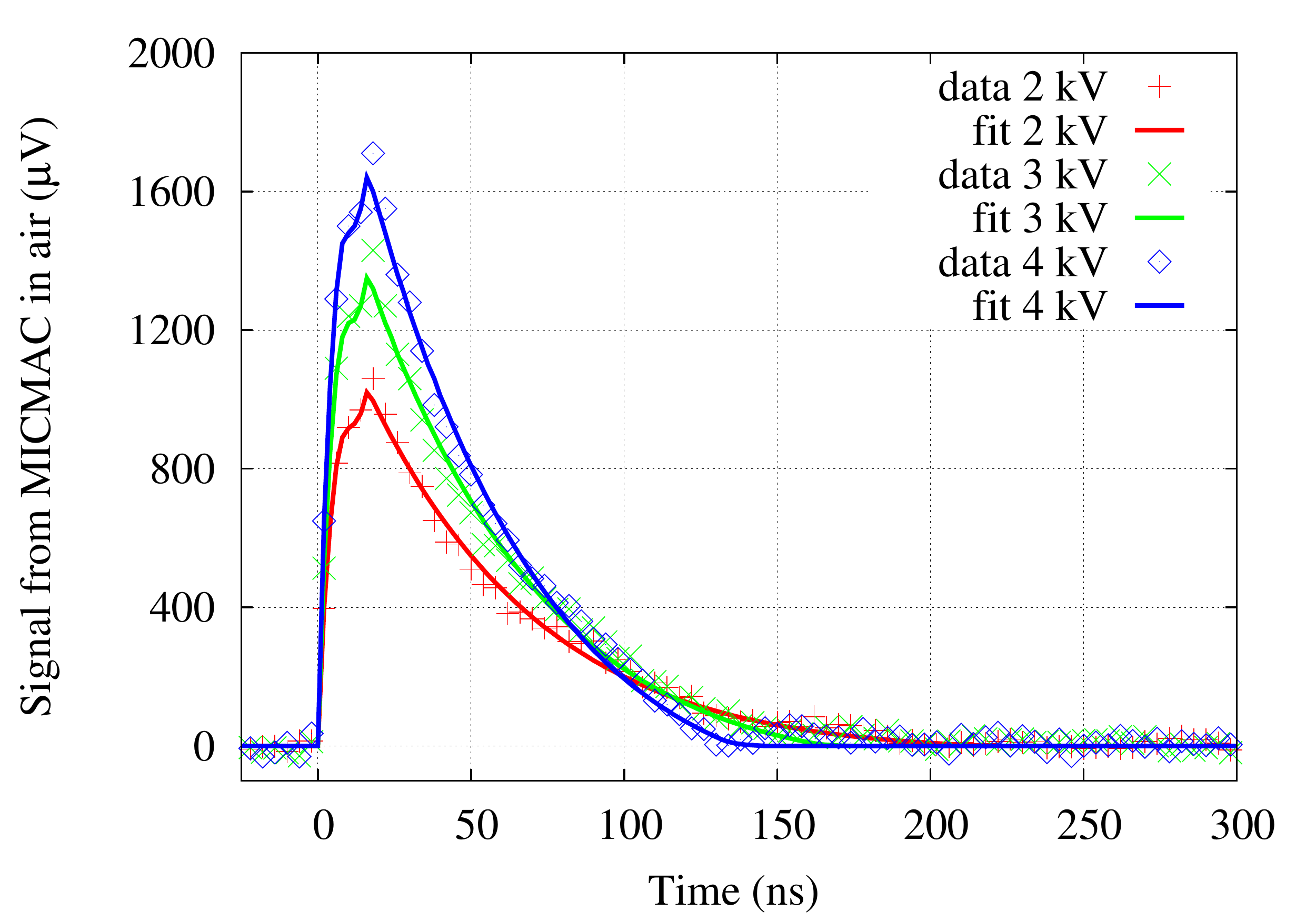}
    \end{center}
    \caption{Measurements and fitted curves for electrons in nitrogen~(a) and air~(b) as a function of time for different applied voltage.}
    \label{MicMac_Ele}
    \end{figure}

    Figure~\ref{MicMac_Ele} presents measurements obtained in nitrogen and air with a 2~ns sampling time. The ion component of the signal is too small to be seen. As explained in Section~\ref{MICMAClowPassFilter}, the need to use a free form filter to model the electronic response can be understood when looking at the unusual but common to every measurement rising shape.
    
    Similarly to ions, the analysis for both air and nitrogen were done fitting all the data at the same time with the same number of ion pairs, the same parameters for the filter but independent drift velocities and attachment time. A subset of electron drift velocities results in air and nitrogen are presented in~Table~\ref{table:MICMAC_EleDriftRes} (full data presented in Annexe~\ref{table:MICMAC_Ele}).
    
    \begin{table}[!ht]
    \caption{MICMAC electron drift velocity and attachment time results ($e^-$ adapted measurements)}\label{table:MICMAC_EleDriftRes}
    \centering
    \footnotesize
    \renewcommand{\arraystretch}{1.3}
    {
    	\begin{tabularx}{\linewidth}{ZZZZZZZZZZZ}
    	\hline
    	\hline
    	HV & $v_{d_{e^-}}^{N_2}$  & $v_{d_{e^-}}^{Air}$ & $T_a$  	 \\
    	\tiny (V)	&  \tiny(mm$\cdot$s$^{-1}$) & \tiny (mm$\cdot$s$^{-1}$) &  \tiny(ns) 	 \\
    	\hline
    	2000 	& 2.43e7				& 2.72e7				& 75.7\\
    	3000 	& 3.27e7				& 3.62e7				& 83.8\\
    	4000 	& 3.99e7				& 4.44e7				& 87.2\\
    	\hline
    	\hline
    	\end{tabularx}
    }
    \end{table}
    
   If cation velocities were really close in both air and nitrogen and seem to depend linearly to the applied voltage, electrons drift velocities was measured about 11~$\%$ higher in air than in nitrogen with MICMAC. When looking at the attachment time are similar than the one measured at the lower sampling rate.
    
    \subsubsection{Comparison between ion and electron measurement.}
    
\begin{figure}[!ht]
    \begin{center}
    \includegraphics[angle=-90, width=0.95\columnwidth]{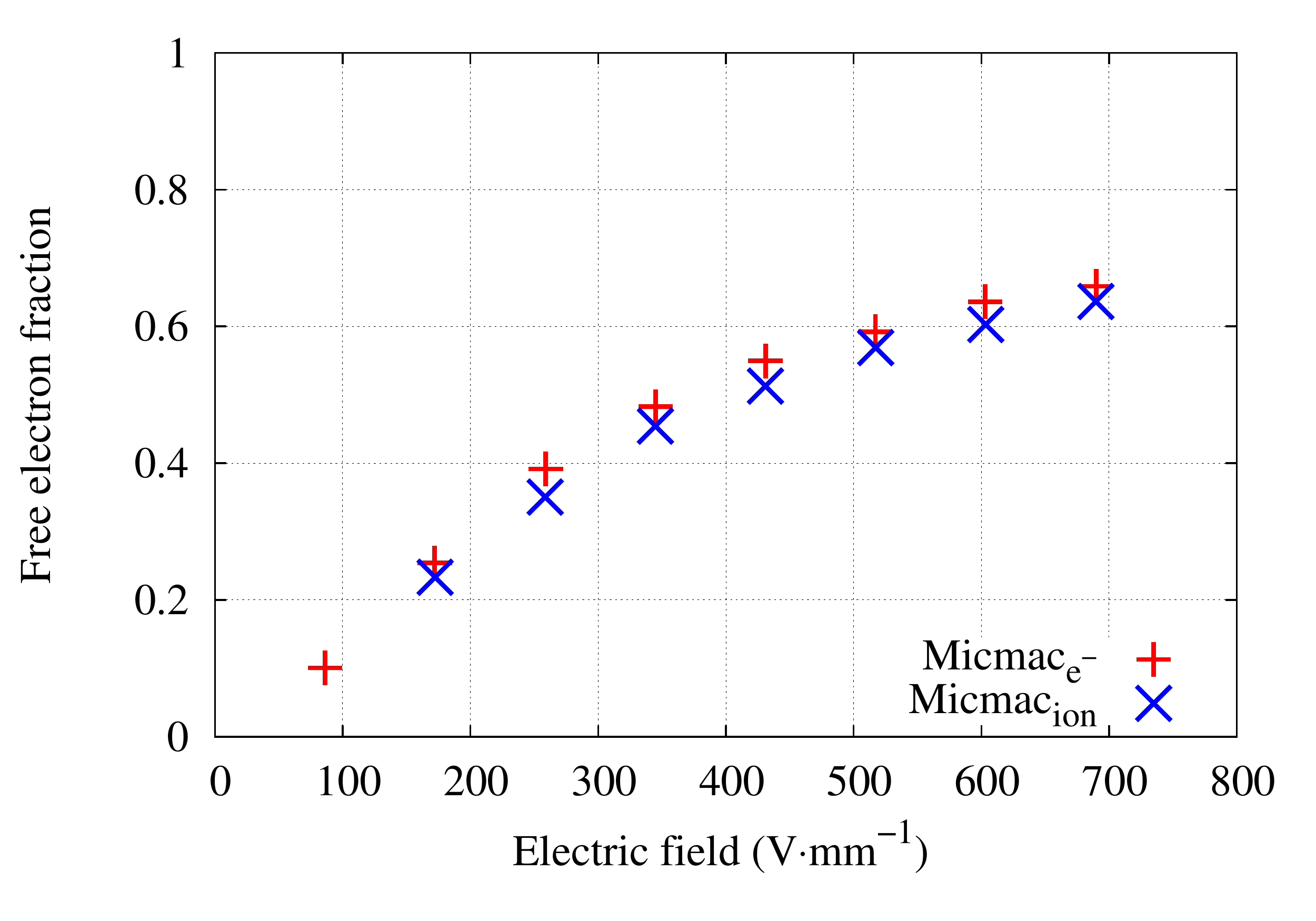}
    \end{center}
    \caption{Free electron fraction using both sets of measurements as a function of the electric field.}
    \label{FreeElectronFraction}
    \end{figure}

    In order to cross validate measurements done using the electron sampling time and those using the ion sampling time, we decided to use the free electron fraction. This quantity, referred in the literature as $p$, is the fraction of the signal-induced by electrons $Q_{e^-}$ (Eq.~\ref{AirIntEle}) over the total signal-induced by negative charges $Q_0$/2. The free electron fraction was measured in both studies independently, through the $T_a$ over $t_{d_{e^-}}$ ratio for ion measurements while both $T_a$ and $t_{d_{e^-}}$ were measured simultaneously in the case of electron measurements. The free electron fraction presents the same evolution compared to electric field values in both studies with a mean difference bellow 7~$\%$ as shown in Fig.~\ref{FreeElectronFraction}.
    
    \subsection{MAGIC ion drift velocity results}

\begin{figure}[!h]
    \begin{center}
    (a)\includegraphics[angle=-90, width=0.95\columnwidth]{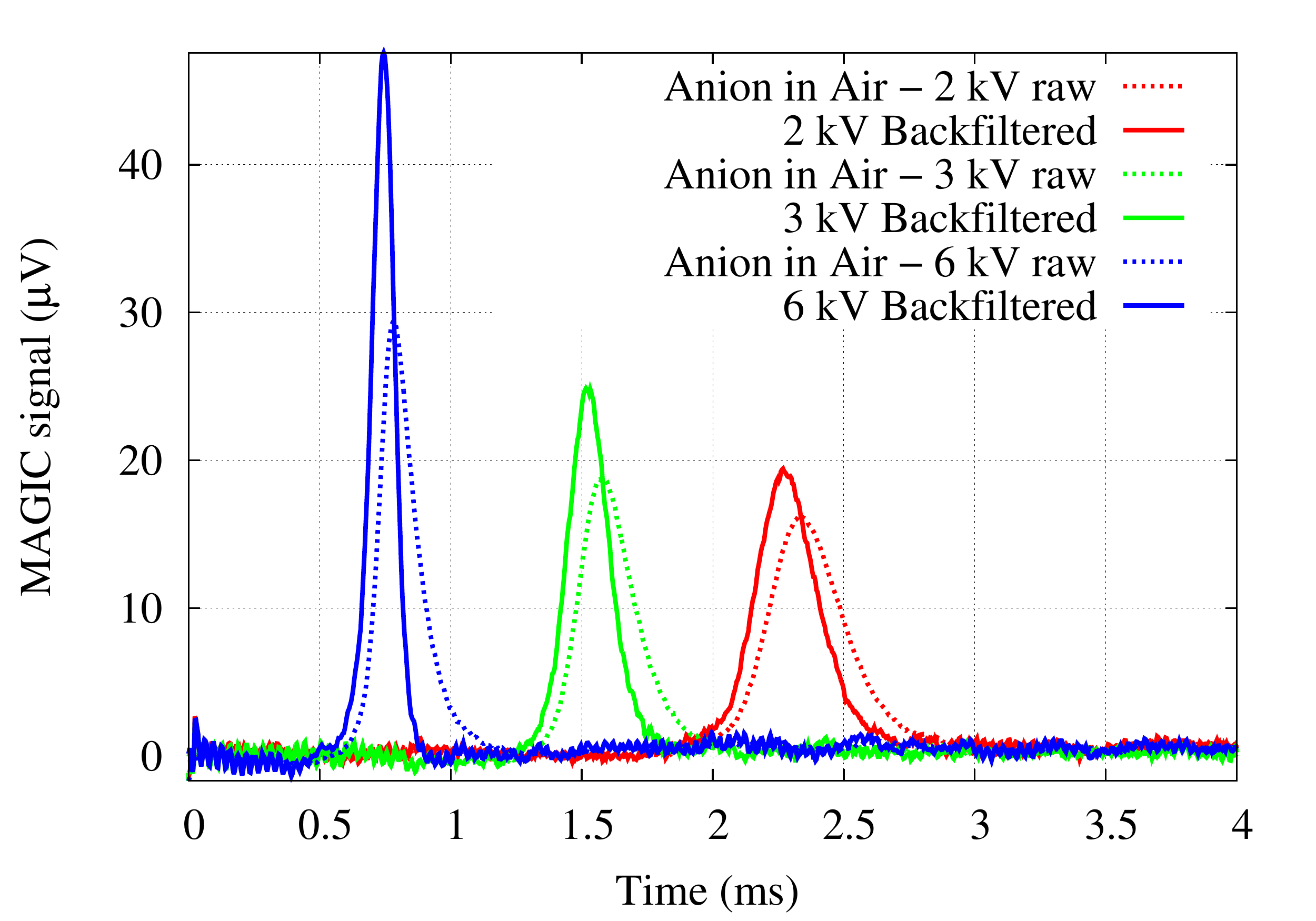}
    (b)\includegraphics[angle=-90, width=0.95\columnwidth]{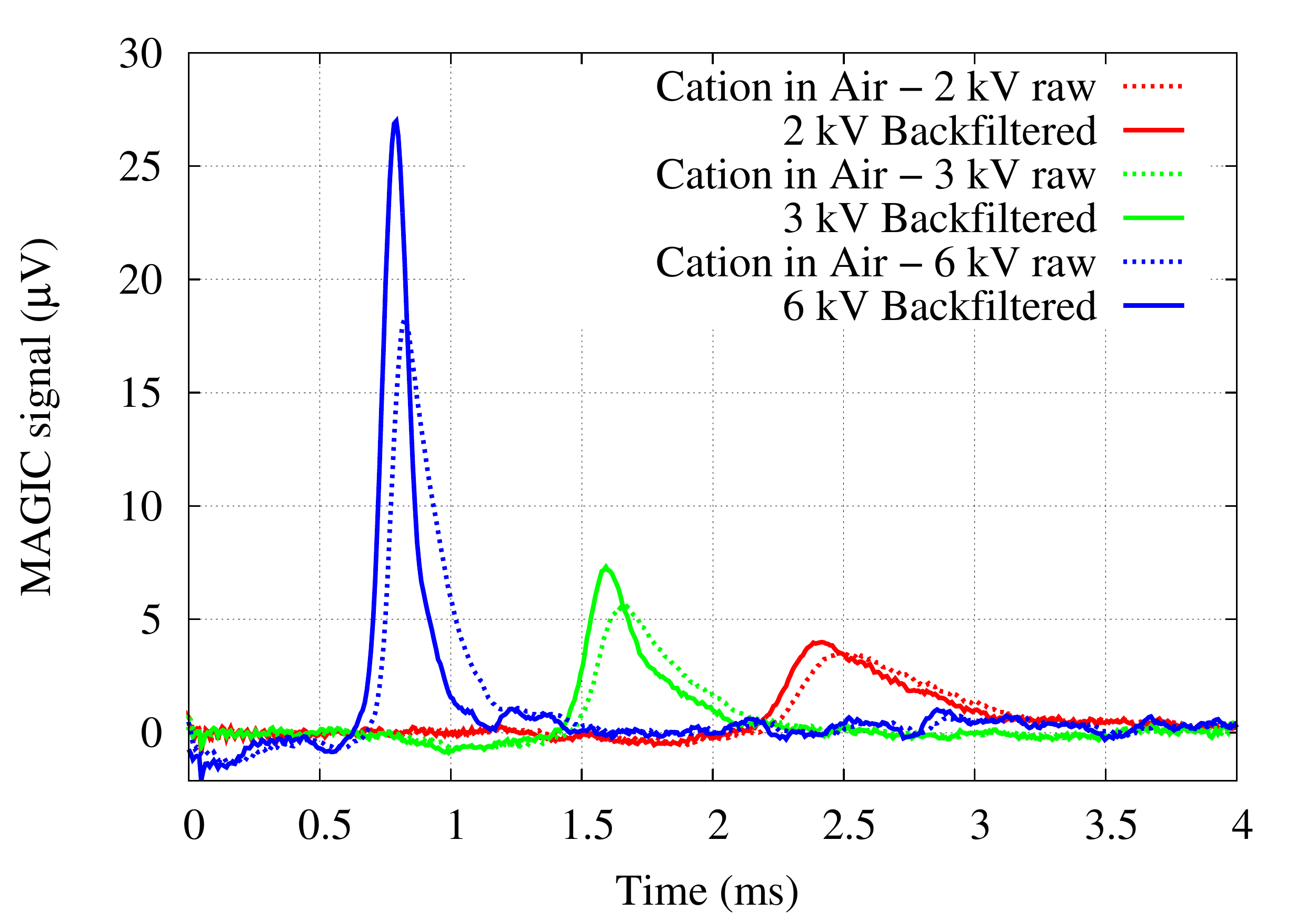}
    \end{center}
    \caption{Measurements and back-filtered curves for negative (a) and positive (b) ions in air with MAGIC as a function of time for different applied voltage.}
    \label{MagicAir}
    \end{figure}

    Time-of-flight signals in MAGIC show different shapes depending on the species studied. Negative ion signals in air tend to keep their symmetrical shape regardless of the applied electric field as shown in Fig.~\ref{MagicAir}~(a) while in the case of positive ions, the decreasing portion seems to widen for decreasing voltage as presented in Fig.~\ref{MagicAir}~(b) from a nearly symmetrical signal at 6~kV and a completely asymmetrical one at 2~kV. Corresponding data are presented in Table~\ref{table:MAGICIonRes} (full data and available in Annexe~\ref{table:MAGIC_Ion}).
    
    This asymmetry, identical in both air and nitrogen for positive ion, might be explained by the fact that positive ions are constituted of a mixture of cations~\cite{Kinetic} with different mobilities.
    
    \begin{table}[!ht]
    \caption{MAGIC ion drift velocity in Air and Nitrogen}
    \centering
    \footnotesize
    \label{table:MAGICIonRes}
    \renewcommand{\arraystretch}{1.3}
    {
    	\begin{tabularx}{\linewidth}{ZZZZZZZZZZZ}
    	\hline
    	\hline
    	HV 	& $v_{d_{+}}^{N_2}$ & $v_{d_{+}}^{Air}$  & $v_{d_{-}}^{Air}$  \\
    	\tiny (V)	&  \tiny(mm$\cdot$s$^{-1}$) & \tiny (mm$\cdot$s$^{-1}$) &  \tiny (mm$\cdot$s$^{-1}$)	 \\
    	\hline
    	2000 	& 1.15e4 				& 1.08e4				& 1.27e4				\\
    	3000 	& 1.77e4				& 1.72e4				& 1.89e4				\\
    	6000 	& 3.54e4				& 3.58e4				& 3.94e4				\\
    	\hline
    	\hline
    	\end{tabularx}
    }
    \end{table}
    
    The ion velocity displays the same linear dependency with the electric field when measured with MAGIC or MICMAC. Cations velocities measured in air and nitrogen are very similar while anions velocities in air was about 10~$\%$ higher.

    \subsection{Comparison to literature}
    
    \subsubsection{Ion drift velocity.}
    
\begin{figure}[!ht]
    \begin{center}
    \includegraphics[angle=-90, width=0.95\columnwidth]{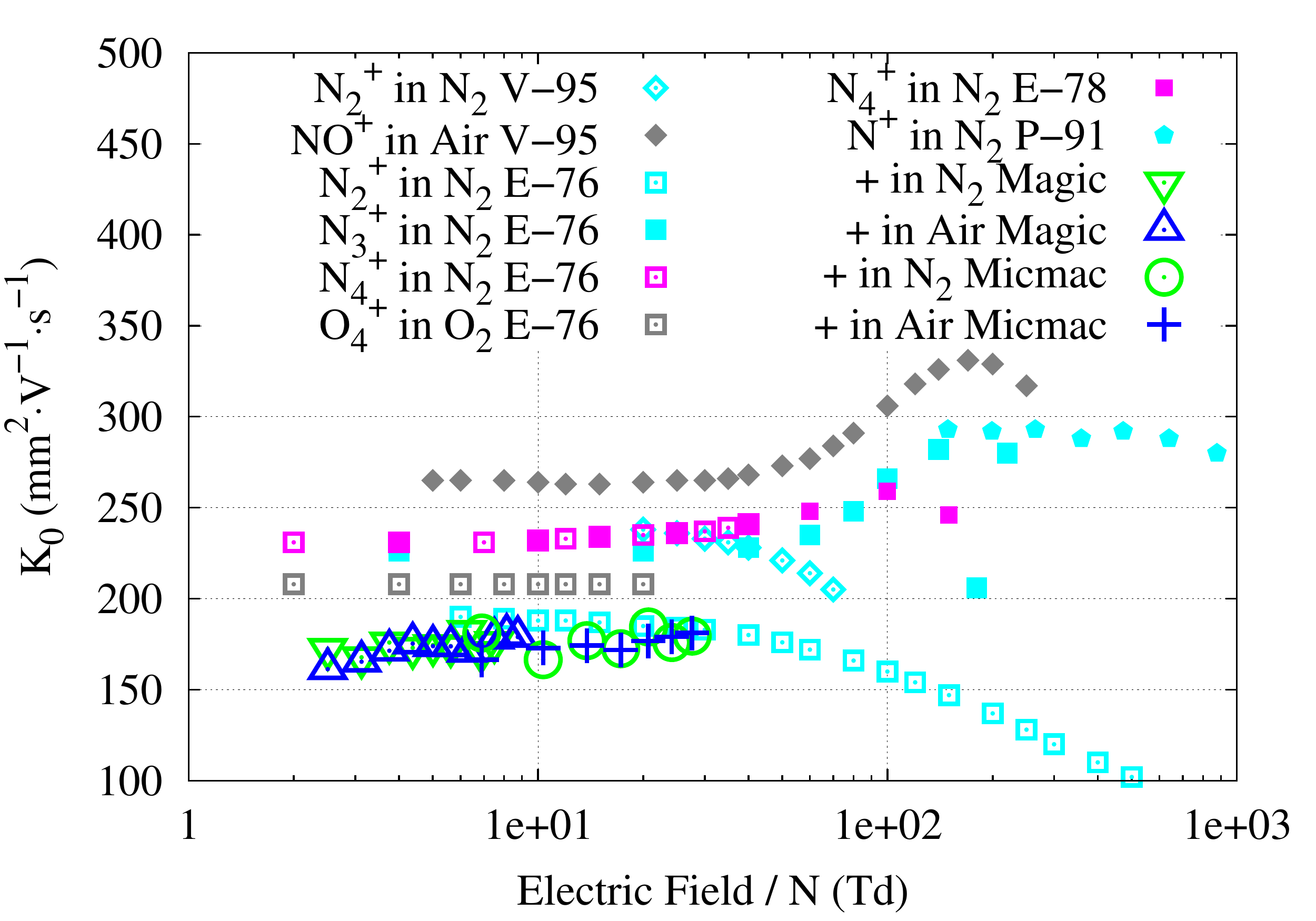}
    \end{center}
    \caption{Standard mobilities of positive ions in nitrogen and air, measurements with MICMAC and MAGIC and along with data from the literature: V-95~\cite{Viehland95}, E-76~\cite{Ellis76}, E-78~\cite{Ellis78} and P-91~\cite{Phelps91}.}
    \label{MobilityN2}
    \end{figure}

\begin{figure}[!ht]
    \begin{center}
    \includegraphics[angle=-90, width=0.95\columnwidth]{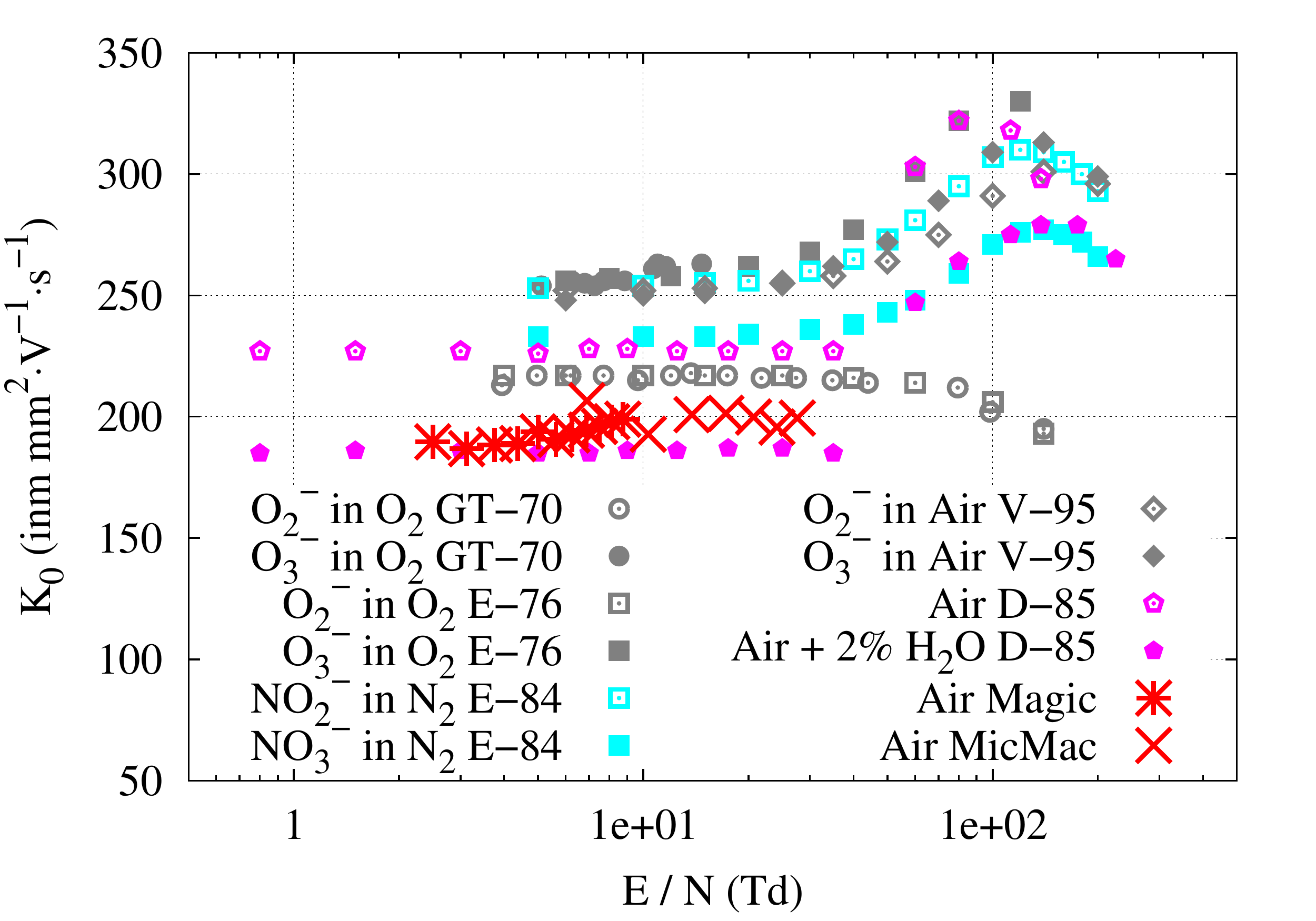}
    \end{center}
    \caption{Reduced mobilities of negative ions in nitrogen and air, measurements with MICMAC and MAGIC and along with data from the literature: GT-70~\cite{gt70}, V-95~\cite{Viehland95}, E-76~\cite{Ellis76}, E-84~\cite{Ellis84} and D-85~\cite{Davies85}.}
    \label{MobilityO2}
    \end{figure}

    Ions velocity measurements with both MICMAC and MAGIC in nitrogen and air are presented in Fig.~\ref{MobilityN2} and \ref{MobilityO2} as standard mobilities ($K_0$, defined in Eq.~\ref{StandardMobility}) along with literature measurements from~\cite{Ellis76,Viehland95,Phelps91,gt70,Ellis84,Davies85}.

    For positive ion measurement, considering the uncertainty on mobilities (1.9\% for MAGIC and 1.5\% for MICMAC, respectively), all measurements performed in this study are in good agreement with themselves as presented in Annexes~\ref{table:MICMAC_Pos} and \ref{table:MAGIC_Ion} for MICMAC and MAGIC, respectively. The standard mobility for cations averaged for both setups is 174~mm$^{2}\cdot$s$^{-1}\cdot$V$^{-1}$($\pm$3.2$\%$) in nitrogen and 175~mm$^{2}\cdot$s$^{-1}\cdot$V$^{-1}$($\pm$2.8$\%$) in the 60 to 700~V$\cdot$mm$^{-1}$ electric field range and are therefore compatible. This indicates that moving species are either $N_2^+$ or $N_4^+$ and are behaving in the same manner in both gases.
    
    The comparison with the literature is difficult as little measurements were performed in air and when they were, the species studied were very different to the "averaged species" measured here. Therefore the most relevant data should be the ones covering either $N_2^+$ or $N_4^+$ in pure nitrogen. Unfortunately, even then, literature data are sparse. As an example, $N_2^+$ in pure $N_2$ measured either by Ellis~(E-76)~\cite{Ellis76} or Viehland~(V-95)~\cite{Viehland95} differed by almost fifteen percent, resulting in the fact that if our measurement could be compatible with $N_2^+$ according to Ellis~\cite{Ellis76}. Overall, their is not any overall agreement with the literature.
    
    Mobility results for anions in air are presented with their uncertainties for MICMAC and MAGIC in Annexes~\ref{table:MICMAC_Neg}~and~\ref{table:MAGIC_Ion}, respectively. The standard mobility for anions averaged for both setups is 195~mm$^{2}\cdot$s$^{-1}\cdot$V$^{-1}$($\pm$2.7$\%$) in the 60 to 700~V$\cdot$mm$^{-1}$ electric field range.
    
    Both MICMAC and MAGIC anion measurements (Fig.~\ref{MobilityO2}) sit between Davies data~\cite{Davies85} in dry air and air containing 2$\%$ $H_20$, that also presents "averaged species" measurements. This tends to indicate that our measurements were done in slightly humid air. However, tests conducted in dry air did not show significant differences. The rest of the subset of literature data presented here proposes higher mobilities values when looking at specific anions in pure $O_2$ and $N_2$ gases.

    \subsubsection{Electron drift velocity.}
    
\begin{figure}[!h]
    \begin{center}
    (a)\includegraphics[angle=-90, width=0.95\columnwidth]{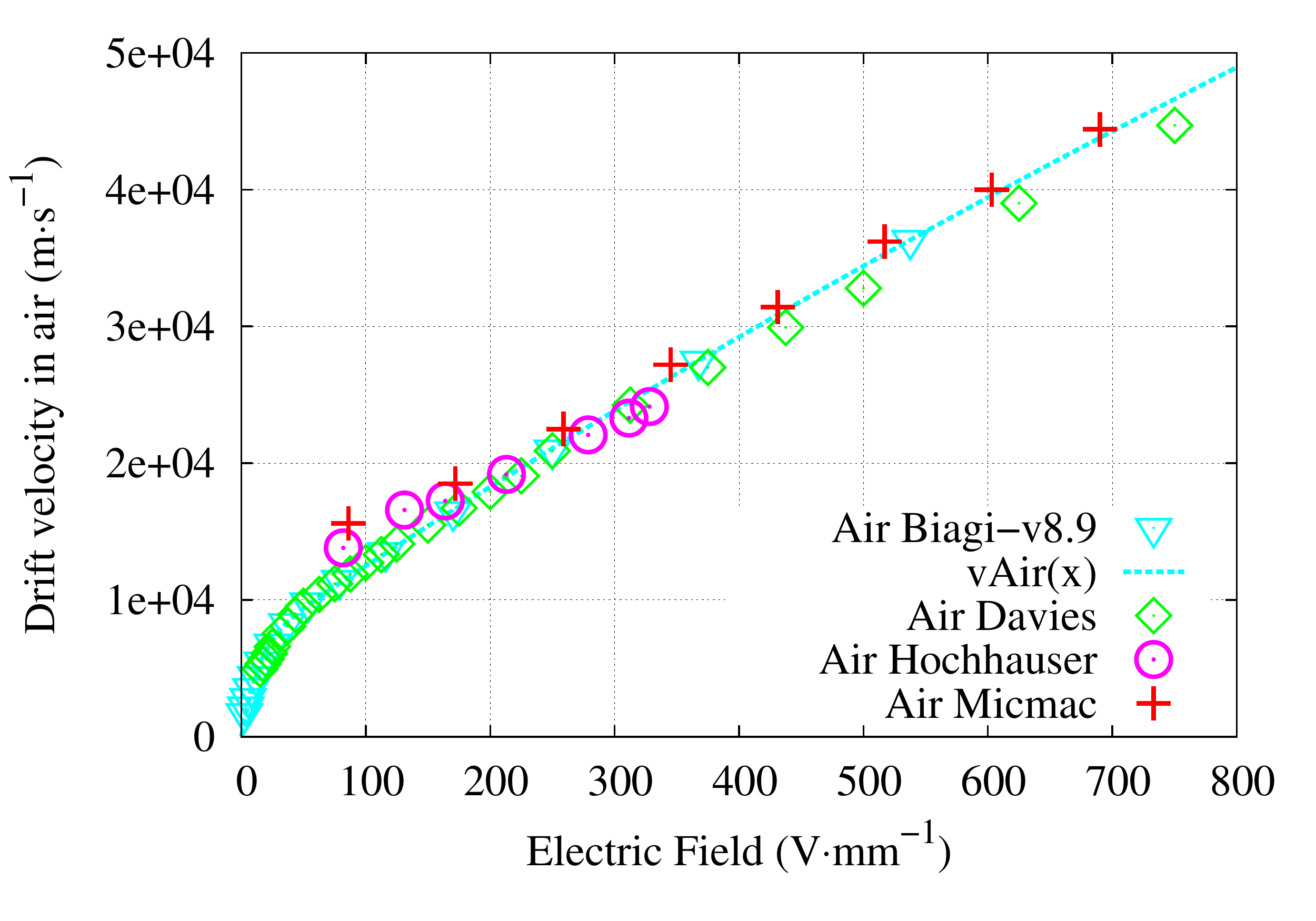}
    (b)\includegraphics[angle=-90, width=0.95\columnwidth]{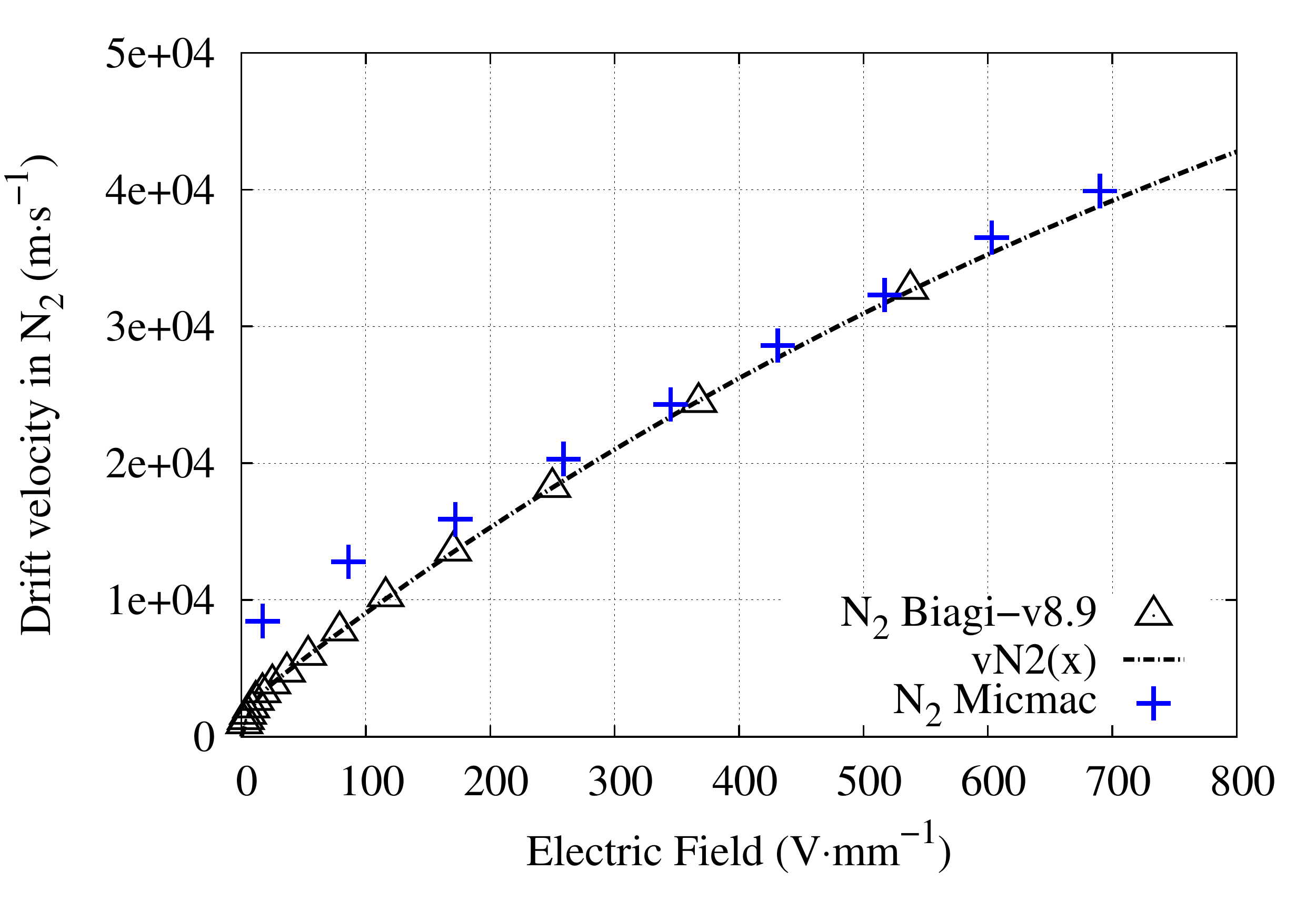}
    \end{center}
    \caption{Drift velocities for electrons in air (a) and nitrogen (b) (in laboratory conditions), measurements with MICMAC along with data from the literature: Hochhauser~\cite{Hochhauser94}, Davies~\cite{Davies85} and Biagi-v8.9 Boltzmann calculations~\cite{Lxcat,Biagi-Roznerski,Biagi-Hagelaar} (vAir and vN2 are functions adjusted on Biagi-v8.9 data).}
    \label{MobilityElectron}
    \end{figure}

     Electron measurements in air and nitrogen are presented in Fig.~\ref{MobilityElectron} displayed as drift velocities. Results are presented along with literature measurements from Hochhauser~\cite{Hochhauser94} and Davies~\cite{Davies85} but also with Biagi-8.9 calculations~\cite{Lxcat,Biagi-Roznerski,Biagi-Hagelaar}. 
     
     MICMAC data in air and nitrogen (full data in Annexe~\ref{table:MICMAC_Ele}) are compatible with literature both in the air and nitrogen in the 200 to 700~V$\cdot$mm$^{-1}$ electric field range but display much higher values at lower electric field.
      
     In air, even if MICMAC values are located above reported measurements, data from Hochhauser tend to display a similar trend compared to the rest of the literature. We currently have no rational to explain why both ours and Hochhauser's measurements diverge from literature at low electric field. Iit is interesting to notice that both measurements were performed with ionization chamber using signal shape analysis. 
     
    \subsubsection{Electronic attachment.}
    
\begin{figure}[!h]
    \begin{center}
    \includegraphics[angle=-90, width=0.95\columnwidth]{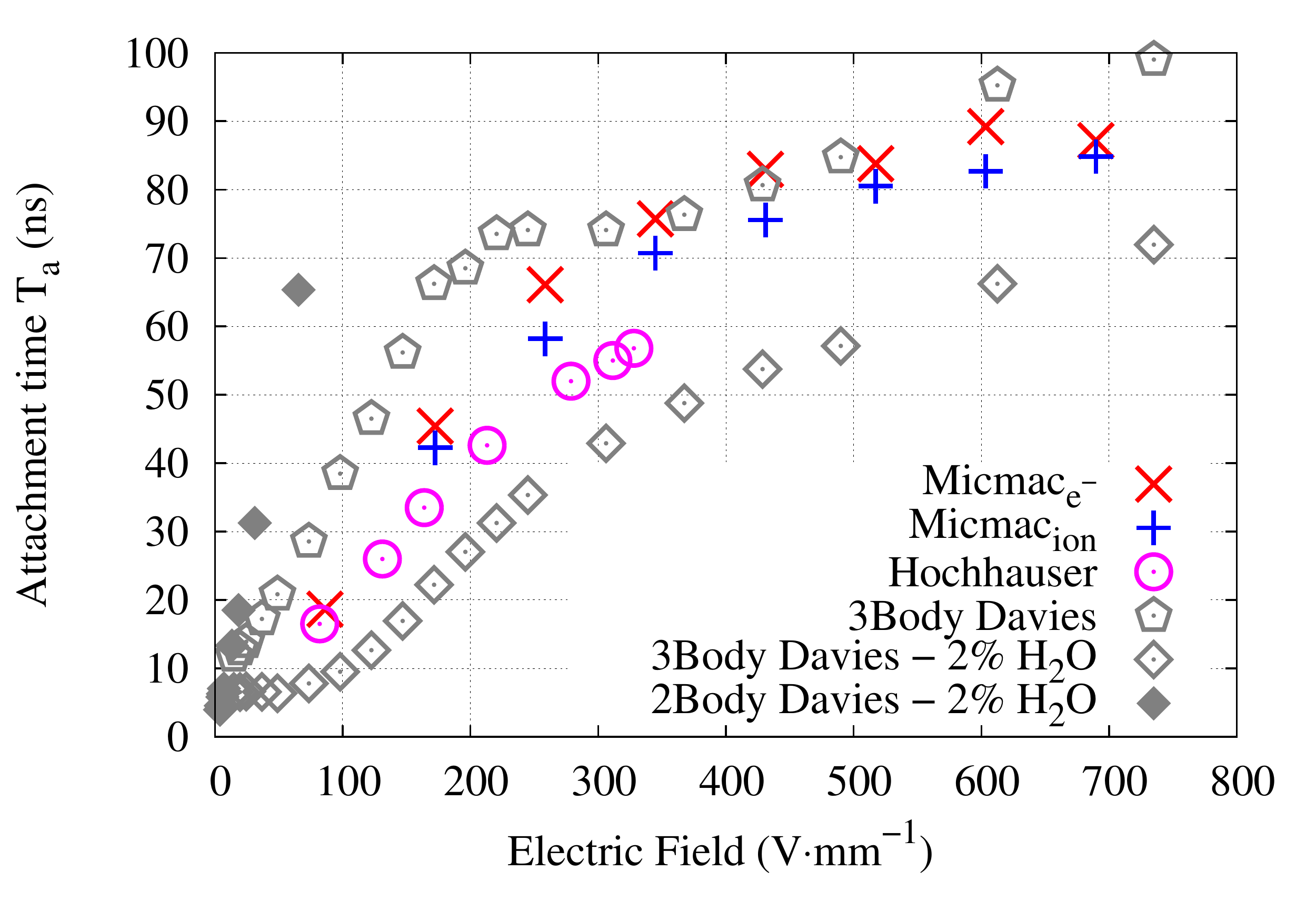}
    \end{center}
    \caption{Attachment time in air measured with MICMAC along with data from the literature: Hochhauser~\cite{Hochhauser94} and Davies~\cite{Davies85}.}
    \label{TdAttachment}
    \end{figure}

    Electronic attachment measurements are presented in Fig.~\ref{TdAttachment} as attachment times with respect to literature data from Hochhauser~\cite{Hochhauser94} and Davies~\cite{Davies85}.
    
    Attachment times were extracted from both electron and ion-adapted time sampling. They display a mean difference of 7.2$\%$ with a very similar behavior over the whole electric field range. It is important to note that, as explained in~\ref{TaOverVeRatio}, attachment times extracted from ion measurements are obtained through the ratio $T_a$ over $t_{d_{e^-}}$ (Eq.~\ref{AirIntEle}). Resulting attachment times values are presented in Annexes~\ref{table:MICMAC_Ele} and~\ref{table:MICMAC_Neg}, for electron measurements and ion measurements, respectively.

    When compared to literature data, our measurements (and Hochhauser's) are positioned between Davies three-body data in dry air and in air with 2$\%$ H$_2$O similarly to what was observed for anion mobilities in air (c.f. Fig.~\ref{MobilityO2}). 
    
    \section{Conclusion and perspectives}
    
    In this study, we aimed to demonstrate the possibility to use actual air ionization chambers to evaluate parameters as the electronic attachment times, electron drift speed and ion mobilities, using  MICMAC detector and to cross validate them (when possible) with a more standard drift tube technic using the detector MAGIC. Those parameters are fundamental when trying to model the response of such detectors. However, there were little data, apart from Hochhausser's measurements~\cite{Hochhauser94,Laitano}, that were performed in conditions adapted to air ionization chambers used in radiotherapy. If electronic drift velocities were well described, only negative ion mobility and attachment times data from Davies~\cite{Davies85}, obtained in a very different context, seem to be relevant for medical use. The very well furnished Ellis and Viehland Tables~\cite{Ellis76,Ellis78,Ellis84,Viehland95} do not provide the effective ion mobilities needed for medical air ionization chambers.
    
    In future work, we will try to use the data of this work to extract recombination values from ionization chambers measurements and to model the efficiency of such detectors in various irradiation conditions. 
    
    \section*{Acknowledgement}
    
    The authors would like to emphasis the role of both IBA and the Region Basse-Normandie for funding this research. We also would like to thank our colleagues at LPC whose expertise allowed us to develop and build the two setups MICMAC and MAGIC.

    \section*{References}
     
\bibliographystyle{elsarticle-num.bst}
    \bibliography{articleMicMac}

\begin{thebibliography}{10}
\expandafter\ifx\csname url\endcsname\relax
  \def\url#1{\texttt{#1}}\fi
\expandafter\ifx\csname urlprefix\endcsname\relax\def\urlprefix{URL }\fi
\expandafter\ifx\csname href\endcsname\relax
  \def\href#1#2{#2} \def\path#1{#1}\fi

\bibitem{IBA}
J.~{Van De Walle}, G.~Boissonnat, Y.~Claereboudt, J.~Colin, J.-M. Fontbonne,
  G.~Krier, D.~Prieels, Dosimetry of pulsed beams in proton therapy, in:
  International Beam Instrumentation Conference, 2014.

\bibitem{Boag}
J.~W. Boag, E.~Hochhauser, O.~A. Balk, The effect of free-electron collection
  on the recombination correction to ionization measurements of pulsed
  radiation, Physics in Medicine and Biology.

\bibitem{IC23}
C.~Courtois, G.~Boissonnat, C.~Brusasco, J.~Colin, D.~Cussol, J.~Fontbonne,
  B.~Marchand, T.~Mertens, S.~de~Neuter, J.~Peronnel, Characterization and
  performances of a monitoring ionization chamber dedicated to iba-universal
  irradiation head for pencil beam scanning, Nuclear Instruments and Methods in
  Physics Research Section A: Accelerators, Spectrometers, Detectors and
  Associated Equipment 736~(0) (2014) 112 -- 117.
\newblock \href
  {http://dx.doi.org/http://dx.doi.org/10.1016/j.nima.2013.10.014}
  {\path{doi:http://dx.doi.org/10.1016/j.nima.2013.10.014}}.

\bibitem{Ellis76}
H.~W. Ellis, R.~Y. Pai, E.~W. McDaniel, Transport propoerties of gaseous ions
  over a wide range, Atomic Data and Nucleair Data Tables 17 (1976) 177.

\bibitem{Kinetic}
I.~A. Kossyi, A.~Y. Kostinsky, A.~A. Matveyev, V.~P. Silakov,
  \href{http://stacks.iop.org/0963-0252/1/i=3/a=011}{Kinetic scheme of the
  non-equilibrium discharge in nitrogen-oxygen mixtures}, Plasma Sources
  Science and Technology 1~(3) (1992) 207.
\newline\urlprefix\url{http://stacks.iop.org/0963-0252/1/i=3/a=011}

\bibitem{Pancheshnyi}
S.~Pancheshnyi,
  \href{http://stacks.iop.org/0022-3727/46/i=15/a=155201}{Effective ionization
  rate in nitrogen-oxygen mixtures}, Journal of Physics D: Applied Physics
  46~(15) (2013) 155201.
\newline\urlprefix\url{http://stacks.iop.org/0022-3727/46/i=15/a=155201}

\bibitem{Ramo}
Z.~He, Review of the shockley-ramo theorem and its application in semiconductor
  gamma-ray detectors, Nuclear Instruments and Methods in Physics Research
  Section A: Accelerators, Spectrometers, Detectors and Associated Equipment
  463~(1--2) (2001) 250 -- 267.
\newblock \href
  {http://dx.doi.org/http://dx.doi.org/10.1016/S0168-9002(01)00223-6}
  {\path{doi:http://dx.doi.org/10.1016/S0168-9002(01)00223-6}}.

\bibitem{refFaster}
\href{faster.in2p3.fr}{Fast acquisition system for nuclear research (lpc caen),
  http://faster.in2p3.fr} [online].

\bibitem{DriftTube}
G.~Eiceman, Z.~Karpas, H.~H. Hill, Ion Mobility Spectrometry, Vol. Third
  Edition, CRC Press, 2013.

\bibitem{FreeFem}
F.~Hecht, \href{http://www.freefem.org/ff++/}{New development in freefem++},
  Journal of Numerical Mathematics 20.
\newline\urlprefix\url{http://www.freefem.org/ff++/}

\bibitem{LeCroy}
LeCroy, Waverunner 6 zi series 400 mhz-- 4 ghz, Tech. rep., Teledyne (2013).

\bibitem{Viehland95}
L.~A. Viehland, E.~Mason, Transport propoerties of gaseous ions over a wide
  range iv, Atomic Data and Nucleair Data Tables 60.

\bibitem{Phelps91}
A.~V. Phelps, Cross sections and swarm coefficients for nitrogen ions and
  neutrals in n2 and argon ions and neutrals in ar for energies from 0.1 ev to
  10 kev, Journal of Physical and Chemical Reference Data.

\bibitem{gt70}
part~of Viehland~database, Raw mobility data from technical reports from
  georgia tech, Tech. rep., Georgia Tech (1970-1974).

\bibitem{Ellis84}
H.~W. Ellis, M.~Thackston, E.~W. McDaniel, E.~A. Mason, Transport propoerties
  of gaseous ions over a wide range iii, Atomic Data and Nucleair Data Tables
  31.

\bibitem{Davies85}
D.~Davies, P.~Chantry, Air chemistry measurements ii, Tech. rep., AIR FORCE
  WEAPONS LABORATORY (1985).

\bibitem{Ellis78}
H.~W. Ellis, E.~W. McDaniel, D.~L. Albritton, L.~A. Viehland, S.~L. Lin, E.~A.
  Mason, Transport propoerties of gaseous ions over a wide range ii, Atomic
  Data and Nucleair Data Tables 22.

\bibitem{Hochhauser94}
E.~Hochhauser, O.~A. Balk, H.~Schneider, W.~Arnold, The significance of the
  lifetime and collection time of free electrons for the recombination
  correction in the ionometric dosimetry of pulsed radiation, Journal of
  Physics D: Applied Physics 27~(3) (1994) 431.
\newblock \href
  {http://dx.doi.org/http://dx.doi.org/10.1088/0022-3727/27/3/001}
  {\path{doi:http://dx.doi.org/10.1088/0022-3727/27/3/001}}.

\bibitem{Lxcat}
\href{http://lxcat.net}{Plasma data exchange project, http://lxcat.net}
  [online].

\bibitem{Biagi-Roznerski}
W.~Roznerski, K.~Leja, Electron drift velocity in hydrogen, nitrogen, oxygen,
  carbon monoxide, carbon dioxide and air at moderate e/n, Journal of Physics
  D: Applied Physics 17.

\bibitem{Biagi-Hagelaar}
G.~J.~M. Hagelaar, L.~C. Pitchford, Solving the boltzmann equation to obtain
  electron transport coefficients and rate coefficients for fluid models,
  Plasma Sources Science and Technology.

\bibitem{Laitano}
R.~Laitano, A.~Guerra, M.~Pimpinella, C.~Caporali, A.~Petrucci,
  \href{http://www.ncbi.nlm.nih.gov/pubmed/17148826}{Charge collection
  efficiency in ionization chambers exposed to electron beams with high dose
  per pulse}, Physics in Medicine and Biology 51~(24) (2006) 6419.
\newline\urlprefix\url{http://www.ncbi.nlm.nih.gov/pubmed/17148826}

\end{thebibliography}


\clearpage

\appendix
    \renewcommand{\thetable}{A\arabic{table}}
    
    \begin{sidewaystable}
    \bigskip\bigskip    \bigskip\bigskip
    \bigskip\bigskip    \bigskip\bigskip
    \bigskip\bigskip    \bigskip\bigskip
    \bigskip\bigskip    \bigskip\bigskip
    \bigskip\bigskip    \bigskip\bigskip
    \bigskip\bigskip    \bigskip\bigskip
    \bigskip\bigskip    \bigskip\bigskip
    \bigskip\bigskip    \bigskip\bigskip
    \bigskip\bigskip    \bigskip\bigskip

    \caption{MICMAC Measurements: Electron mobilities and electronic attachment times in Nitrogen and Air (at 291$\pm$2.9~K and 1015$\pm$8.6~hPa)} \label{table:MICMAC_Ele}
    \footnotesize
    \renewcommand{\arraystretch}{1.3}{
    \begin{tabularx}{0.97\textwidth}{cZZcccccccccccccc}
    					\hline
    					\hline
     HV & E   & $t_{d_{e^-}}^{N_2}$  & $v_{d_{e^-}}^{N_2}$  &  $K_{e^-}^{N_2}$ &  $K_{0_{e^-}}^{N_2}$   & $t_{d_{e^-}}^{Air}$ & $v_{d_{e^-}}^{Air}$ 	& $K_{e^-}^{Air}$	&  $K_{0_{e^-}}^{Air}$	& Ta \\
    
    					\tiny	(V)& \tiny (V$\cdot$mm$^{-1}$) & \tiny(s) & \tiny (mm$\cdot$s$^{-1}$)& \tiny(mm$^{2}\cdot$s$^{-1}\cdot$V$^{-1}$)	& \tiny(mm$^2$$\cdot$s$^{-1}\cdot$V$^{-1}$)  & \tiny (s) & \tiny(mm$\cdot$s$^{-1}$)	& \tiny(mm$^{2}\cdot$s$^{-1}\cdot$V$^{-1}$) & \tiny(mm$^2$$\cdot$s$^{-1}\cdot$V$^{-1}$)  &  \tiny (s)  	\\
    					\hline
     100		& 17.2 {\tiny$\pm$0.58$\%$}	& 6.86e-7  {\tiny$\pm$0.01$\%$}		& 8.46e6 {\tiny$\pm$0.50$\%$}		& 4.91e5 {\tiny$\pm$0.76$\%$}  	& 4.58e5 {\tiny$\pm$1.51$\%$}  \\
     500		& 86.2 {\tiny$\pm$0.50$\%$}	& 4.52e-7 {\tiny$\pm$0.01$\%$}		& 1.28e7 {\tiny$\pm$0.50$\%$}		& 1.49e5 {\tiny$\pm$0.71$\%$}  	& 1.39e5 {\tiny$\pm$1.48$\%$}  
    		& 3.72e-7 {\tiny$\pm$0.05$\%$}		& 1.56e7 {\tiny$\pm$0.50$\%$} 		& 1.81e5 {\tiny$\pm$0.71$\%$}  				& 1.69e5 {\tiny$\pm$1.48$\%$}  &	18.6 {\tiny$\pm$1.40$\%$} \\
    1000		& 172 {\tiny$\pm$0.50$\%$}	& 3.64e-7 {\tiny$\pm$0.01$\%$}		& 1.59e7 {\tiny$\pm$0.50$\%$}		& 9.24e4 {\tiny$\pm$0.70$\%$}  	& 8.63e4 {\tiny$\pm$1.48$\%$}  
    																	& 3.14e-7 {\tiny$\pm$0.03$\%$}		& 1.85e7 {\tiny$\pm$0.50$\%$}		& 1.07e4 {\tiny$\pm$0.71$\%$}  				& 1.00e5 {\tiny$\pm$1.48$\%$}  &	45.4 {\tiny$\pm$0.35$\%$}\\
    1500		& 259 {\tiny$\pm$0.50$\%$}	& 2.85e-7 {\tiny$\pm$0.01$\%$}		& 2.03e7 {\tiny$\pm$0.50$\%$}		& 7.86e4 {\tiny$\pm$0.70$\%$}  	& 7.34e4 {\tiny$\pm$1.48$\%$}  
    																	& 2.58e-7 {\tiny$\pm$0.03$\%$}		& 2.25e7 {\tiny$\pm$0.50$\%$}		& 8.69e4 {\tiny$\pm$0.70$\%$}  				& 8.12e4 {\tiny$\pm$1.48$\%$}  &	66.1 {\tiny$\pm$0.19$\%$}\\
    2000		& 345 {\tiny$\pm$0.50$\%$}	& 2.38e-7 {\tiny$\pm$0.01$\%$}		& 2.43e7 {\tiny$\pm$0.50$\%$}		& 7.05e4 {\tiny$\pm$0.70$\%$}  	& 6.59e4 {\tiny$\pm$1.48$\%$}
    																	& 2.13e-7 {\tiny$\pm$0.03$\%$}		& 2.72e7 {\tiny$\pm$0.50$\%$}		& 7.89e4 {\tiny$\pm$0.70$\%$}  				& 7.37e4 {\tiny$\pm$1.48$\%$}  &	75.7 {\tiny$\pm$0.16$\%$}\\
    2500		& 431 {\tiny$\pm$0.50$\%$}	& 2.03e-7 {\tiny$\pm$0.01$\%$}		& 2.86e7 {\tiny$\pm$0.50$\%$}		& 6.63e4 {\tiny$\pm$0.70$\%$}  	& 6.19e4 {\tiny$\pm$1.48$\%$}  
    																	& 1.85e-7 {\tiny$\pm$0.03$\%$}		& 3.14e7 {\tiny$\pm$0.50$\%$}		& 7.29e4 {\tiny$\pm$0.70$\%$}  				& 6.81e4 {\tiny$\pm$1.48$\%$}  &	83.0 {\tiny$\pm$0.13$\%$}\\
    3000		& 517 {\tiny$\pm$0.50$\%$}	& 1.80e-7 {\tiny$\pm$0.02$\%$}		& 3.27e7 {\tiny$\pm$0.50$\%$}		& 6.25e4 {\tiny$\pm$0.70$\%$}  	& 5.83e4 {\tiny$\pm$1.48$\%$}
    																	& 1.60e-7 {\tiny$\pm$0.03$\%$}		& 3.62e7 {\tiny$\pm$0.50$\%$}		& 7.00e4 {\tiny$\pm$0.70$\%$}  				& 6.54e4 {\tiny$\pm$1.48$\%$}  &	83.8 {\tiny$\pm$0.13$\%$}\\
    3500		& 603 {\tiny$\pm$0.50$\%$}	& 1.59e-7 {\tiny$\pm$0.02$\%$}		& 3.65e7 {\tiny$\pm$0.50$\%$}	& 6.05e4 {\tiny$\pm$0.70$\%$}  	& 5.65e4 {\tiny$\pm$1.48$\%$} 
    																	& 1.45e-7 {\tiny$\pm$0.03$\%$}		& 4.00e7 {\tiny$\pm$0.50$\%$} 		& 6.62e4 {\tiny$\pm$0.70$\%$}  				& 6.19e4 {\tiny$\pm$1.48$\%$}  &	89.2 {\tiny$\pm$0.12$\%$}\\
    4000		& 690 {\tiny$\pm$0.50$\%$}	& 1.45e-7 {\tiny$\pm$0.02$\%$}		& 3.99e7 {\tiny$\pm$0.50$\%$}		& 5.79e4 {\tiny$\pm$0.70$\%$}  	& 5.40e4 {\tiny$\pm$1.48$\%$}  
    																	& 1.31e-7 {\tiny$\pm$0.04$\%$}		& 4.44e7 {\tiny$\pm$0.50$\%$}		& 6.44e4 {\tiny$\pm$0.70$\%$}  				& 6.02e4 {\tiny$\pm$1.48$\%$}  &	87.2  {\tiny$\pm$0.14$\%$}\\
    					\hline
    					\hline
    \end{tabularx}
    
    \bigskip\bigskip
    \bigskip\bigskip
    
    \caption{MICMAC Measurements: Cation drift velocities and mobilities in Nitrogen and Air (at 291$\pm$2.9~K and 1015$\pm$8.6~hPa)}\label{table:MICMAC_Pos}
    \begin{tabularx}{0.97\linewidth}{ccccccccccccc}
    					\hline
    					\hline
    	HV 				& E& $t_{d_{+}}^{N_2}$		& $v_{d_{+}}^{N_2}$ 
    																				& $K_{+}^{N_2}$ 
    																								   & $K_{0_+}^{N_2}$ 
    																								   					& $t_{d_{+}}^{Air}$					
    																																	& $v_{d_{+}}^{Air}$ 		
    																																						& $K_{+}^{Air}$																																														&  $K_{0_+}^{Air}$  \\
    \tiny	(V)& \tiny (V$\cdot$mm$^{-1}$) & \tiny(s) & \tiny (mm$\cdot$s$^{-1}$)& \tiny(mm$^{2}\cdot$s$^{-1}\cdot$V$^{-1}$)	& \tiny(mm$^2$$\cdot$s$^{-1}\cdot$V$^{-1}$)  & \tiny (s) & \tiny(mm$\cdot$s$^{-1}$)	& \tiny(mm$^{2}\cdot$s$^{-1}\cdot$V$^{-1}$) & \tiny(mm$^2$$\cdot$s$^{-1}\cdot$V$^{-1}$)  	\\
    					\hline
    1000		& 172 {\tiny$\pm$0.50$\%$}	& 1.73e-4 {\tiny$\pm$0.13$\%$}	& 3.34e4 {\tiny$\pm$0.52$\%$}	& 194 {\tiny$\pm$0.72$\%$}  & 181 {\tiny$\pm$1.49$\%$}  & 1.89e-4 {\tiny$\pm$0.10$\%$}	& 3.07e4 {\tiny$\pm$0.51$\%$}	& 178 {\tiny$\pm$0.71$\%$}  & 166 {\tiny$\pm$1.48$\%$}  \\
    1500		& 259 {\tiny$\pm$0.50$\%$}	& 1.26e-4 {\tiny$\pm$0.16$\%$}	& 4.61e4 {\tiny$\pm$0.52$\%$}	& 178 {\tiny$\pm$0.72$\%$}  & 167 {\tiny$\pm$1.49$\%$}  & 1.21e-4 {\tiny$\pm$0.12$\%$}	& 4.79e4 {\tiny$\pm$0.51$\%$}	& 185 {\tiny$\pm$0.71$\%$}  & 173 {\tiny$\pm$1.49$\%$}  \\
    2000		& 345 {\tiny$\pm$0.50$\%$}	& 8.88e-5 {\tiny$\pm$0.18$\%$}	& 6.53e4 {\tiny$\pm$0.53$\%$}	& 189 {\tiny$\pm$0.73$\%$}  & 177 {\tiny$\pm$1.49$\%$}  & 9.02e-5 {\tiny$\pm$0.14$\%$}	& 6.43e4 {\tiny$\pm$0.52$\%$}	& 187 {\tiny$\pm$0.72$\%$}  & 174 {\tiny$\pm$1.49$\%$}  \\
    2500		& 431 {\tiny$\pm$0.50$\%$}	& 7.28e-5 {\tiny$\pm$0.20$\%$}	& 7.96e4 {\tiny$\pm$0.53$\%$}	& 185 {\tiny$\pm$0.73$\%$}  & 173 {\tiny$\pm$1.49$\%$}  & 7.31e-5 {\tiny$\pm$0.17$\%$}	& 7.93e4 {\tiny$\pm$0.53$\%$}	& 183 {\tiny$\pm$0.72$\%$}  & 172 {\tiny$\pm$1.49$\%$}  \\
    3000		& 517 {\tiny$\pm$0.50$\%$}	& 5.68e-5 {\tiny$\pm$0.24$\%$}	& 1.02e5 {\tiny$\pm$0.55$\%$}	& 197 {\tiny$\pm$0.74$\%$}  & 184 {\tiny$\pm$1.50$\%$}  & 5.92e-5 {\tiny$\pm$0.18$\%$}	& 9.79e4 {\tiny$\pm$0.53$\%$}	& 189 {\tiny$\pm$0.73$\%$}  & 177 {\tiny$\pm$1.49$\%$}  \\
    3500		& 603 {\tiny$\pm$0.50$\%$}	& 5.10e-5 {\tiny$\pm$0.25$\%$}	& 1.14e5 {\tiny$\pm$0.56$\%$}	& 188 {\tiny$\pm$0.75$\%$}  & 176 {\tiny$\pm$1.50$\%$}  & 5.01e-5 {\tiny$\pm$0.20$\%$}	& 1.16e5 {\tiny$\pm$0.53$\%$}	& 192 {\tiny$\pm$0.73$\%$}  & 179 {\tiny$\pm$1.49$\%$}  \\
    4000		& 690 {\tiny$\pm$0.50$\%$}	& 4.38e-5 {\tiny$\pm$0.27$\%$}	& 1.33e5 {\tiny$\pm$0.57$\%$}	& 192 {\tiny$\pm$0.75$\%$}  & 179 {\tiny$\pm$1.51$\%$}  & 4.34e-5 {\tiny$\pm$0.26$\%$}	& 1.34e5 {\tiny$\pm$0.56$\%$}	& 194 {\tiny$\pm$0.75$\%$}  & 181 {\tiny$\pm$1.50$\%$}  \\
    					\hline
    					\hline
    \end{tabularx}}
    \end{sidewaystable}


\clearpage
    \begin{sidewaystable}

    \renewcommand{\arraystretch}{1.3}{
    \bigskip\bigskip    \bigskip\bigskip
    \bigskip\bigskip    \bigskip\bigskip
    \bigskip\bigskip    \bigskip\bigskip
    \bigskip\bigskip    \bigskip\bigskip
    \bigskip\bigskip    \bigskip\bigskip
    \bigskip\bigskip    \bigskip\bigskip
    \bigskip\bigskip    \bigskip\bigskip
    \bigskip\bigskip    \bigskip\bigskip
    \bigskip\bigskip    \bigskip\bigskip
    
    \footnotesize
    
    \caption{MICMAC Measurements: Anion drift velocities, mobilities and attachment time in Air (at 291$\pm$2.9~K and 1015$\pm$8.6~hPa)}\label{table:MICMAC_Neg}
    \begin{tabularx}{0.97\linewidth}{ZZZZZZZ}
    					\hline
    					\hline
    HV 	& E 	& $t_{d_{-}}$	& $v_{d_{-}}$ 	& $K_{-}$	&  $K{_0}{^{-}}$ 	& Ta  \\
    \tiny	(V)& \tiny (V$\cdot$mm$^{-1}$) & \tiny(s) & \tiny (mm$\cdot$s$^{-1}$)& \tiny(mm$^{2}\cdot$s$^{-1}\cdot$V$^{-1}$)	& \tiny(mm$^2$$\cdot$s$^{-1}\cdot$V$^{-1}$)  & \tiny (s)	\\
    					\hline1000		& 172 {\tiny$\pm$0.50$\%$}	& 1.52e-4 {\tiny$\pm$0.13$\%$}			& 3.81e4 {\tiny$\pm$0.52$\%$}			& 221 {\tiny $\pm$0.72$\%$}  				& 207 {\tiny $\pm$1.49$\%$} 	&	42.3 {\tiny $\pm$0.58$\%$} \\
    1500		& 259 {\tiny$\pm$0.50$\%$}	& 1.09e-4 {\tiny$\pm$0.21$\%$}			& 5.34e4 {\tiny $\pm$0.54$\%$}			& 206 {\tiny $\pm$0.73$\%$}  				& 193 {\tiny $\pm$1.50$\%$} 	&	58.2 {\tiny $\pm$0.53$\%$} \\
    2000		& 345 {\tiny$\pm$0.50$\%$}	& 7.82e-5 {\tiny$\pm$0.27$\%$}			& 7.42e4 {\tiny $\pm$0.57$\%$}			& 215 {\tiny $\pm$0.75$\%$}  				& 201 {\tiny $\pm$1.51$\%$} 	&	70.7 {\tiny $\pm$0.52$\%$} \\
    2500		& 431 {\tiny$\pm$0.50$\%$}	& 6.24e-5 {\tiny$\pm$0.35$\%$}			& 9.30e4 {\tiny $\pm$0.61$\%$}			& 216 {\tiny $\pm$0.79$\%$}  				& 201 {\tiny $\pm$1.52$\%$} 	&	75.6 {\tiny $\pm$0.52$\%$} \\
    3000		& 517 {\tiny$\pm$0.50$\%$}	& 5.24e-5 {\tiny$\pm$0.43$\%$}			& 1.11e5 {\tiny $\pm$0.66$\%$}			& 214 {\tiny $\pm$0.83$\%$}  				& 200 {\tiny $\pm$1.54$\%$} 	&	80.5 {\tiny $\pm$0.52$\%$} \\
    3500		& 603 {\tiny$\pm$0.50$\%$}	& 4.58e-5 {\tiny$\pm$0.48$\%$}			& 1.27e5 {\tiny $\pm$0.69$\%$}			& 210 {\tiny $\pm$0.85$\%$}  				& 196 {\tiny $\pm$1.56$\%$} 	&	82.7 {\tiny $\pm$0.51$\%$} \\
    4000		& 690 {\tiny$\pm$0.50$\%$}	& 3.94e-5 {\tiny$\pm$0.73$\%$}			& 1.47e5 {\tiny $\pm$0.88$\%$}			& 214 {\tiny $\pm$1.01$\%$}  				& 199 {\tiny $\pm$1.65$\%$} 	&	84.8 {\tiny $\pm$0.53$\%$} \\
    					\hline
    					\hline
    \end{tabularx}
    
    \bigskip\bigskip
    \bigskip\bigskip

    \caption{MAGIC Measurements: Ion drift velocities and mobilities in Nitrogen and Air (at 291$\pm$2.9~K and 1015$\pm$8.6~hPa)}\label{table:MAGIC_Ion}
    \begin{tabularx}{0.97\linewidth}{ZZZZZZZZZZZZZZ}
    					\hline
    					\hline
    HV 		& E	& $t_{d_{+}}^{N_2}$	& $v_{d_{+}}^{N_2}$   & $K_{+}^{N_2}$ 	 & $K_{0_+}^{N_2}$  & $t_{d_{+}}^{Air}$			& $v_{d_{+}}^{Air}$ 	 	& $K_{+}^{Air}$		&  $K_{0_+}^{Air}$ & $t_{d_{-}}^{Air}$ & $v_{d_{-}}^{Air}$ 		& $K_{-}^{Air}$			&  $K_{0_-}^{Air}$\\
    \tiny	(V)& \tiny (V$\cdot$mm$^{-1}$) & \tiny(s) & \tiny (mm$\cdot$s$^{-1}$)& \tiny (mm$^{2}$ $\cdot$s$^{-1}\cdot$V$^{-1}$)	& \tiny (mm$^{2}$ $\cdot$s$^{-1}\cdot$V$^{-1}$)	  & \tiny (s) & \tiny(mm$\cdot$s$^{-1}$)	& \tiny (mm$^{2}$ $\cdot$s$^{-1}\cdot$V$^{-1}$)	  & \tiny (mm$^{2}$ $\cdot$s$^{-1}\cdot$V$^{-1}$)	 & \tiny (s) & \tiny(mm$\cdot$s$^{-1}$)	&\tiny (mm$^{2}$ $\cdot$s$^{-1}\cdot$V$^{-1}$)	 &\tiny (mm$^{2}$ $\cdot$s$^{-1}\cdot$V$^{-1}$)		\\
    					\hline
    2000 	& 62.5 {\tiny$\pm$0.9$\%$}	& 2.51e-3		& 1.15e4 {\tiny$\pm$1.0$\%$}	& 185 {\tiny$\pm$1.34$\%$}  & 172 {\tiny$\pm$1.88$\%$}  & 2.69e-3	& 1.08e4 {\tiny$\pm$1.0$\%$}	& 173 {\tiny$\pm$1.34$\%$}  & 161 {\tiny$\pm$1.88$\%$} & 2.28e-3	& 1.27e4 {\tiny$\pm$1.0$\%$}	& 203 {\tiny$\pm$1.34$\%$}  & 190 {\tiny$\pm$1.88$\%$}  \\
    2500 	& 78.1 {\tiny$\pm$0.9$\%$}	& 2.07e-3		& 1.40e4 {\tiny$\pm$1.0$\%$}	& 180 {\tiny$\pm$1.34$\%$}  & 168 {\tiny$\pm$1.88$\%$}  & 2.10e-3	& 1.38e4 {\tiny$\pm$1.0$\%$}	& 177 {\tiny$\pm$1.34$\%$}  & 165 {\tiny$\pm$1.88$\%$} & 1.86e-3	& 1.56e4 {\tiny$\pm$1.0$\%$}	& 200 {\tiny$\pm$1.34$\%$}  & 187 {\tiny$\pm$1.88$\%$}  \\
    3000 	& 93.8 {\tiny$\pm$0.9$\%$}	& 1.64e-3		& 1.77e4 {\tiny$\pm$1.0$\%$}	& 188 {\tiny$\pm$1.34$\%$}  & 176 {\tiny$\pm$1.88$\%$}  & 1.69e-3	& 1.72e4 {\tiny$\pm$1.0$\%$}	& 184 {\tiny$\pm$1.34$\%$}  & 171 {\tiny$\pm$1.88$\%$}  & 1.53e-3	& 1.89e4 {\tiny$\pm$1.0$\%$}	& 202 {\tiny$\pm$1.34$\%$}  & 188 {\tiny$\pm$1.88$\%$}  \\
    3500 	& 109 {\tiny$\pm$0.9$\%$}	& 1.43e-3		& 2.03e4 {\tiny$\pm$1.00$\%$}	& 185 {\tiny$\pm$1.34$\%$}  & 173 {\tiny$\pm$1.88$\%$}  & 1.41e-3	& 2.05e4 {\tiny$\pm$1.0$\%$}	& 188 {\tiny$\pm$1.34$\%$}  & 175 {\tiny$\pm$1.88$\%$}  & 1.31e-3	& 2.21e4 {\tiny$\pm$1.0$\%$}	& 202 {\tiny$\pm$1.34$\%$}  & 189 {\tiny$\pm$1.88$\%$}  \\
    4000 	& 125 {\tiny$\pm$0.9$\%$}	& 1.24e-3		& 2.34e4 {\tiny$\pm$1.0$\%$}	& 187 {\tiny$\pm$1.34$\%$}  & 175 {\tiny$\pm$1.88$\%$}  & 1.25e-3	& 2.33e4 {\tiny$\pm$1.0$\%$}	& 186 {\tiny$\pm$1.34$\%$}  & 174 {\tiny$\pm$1.88$\%$}  & 1.12e-3	& 2.59e4 {\tiny$\pm$1.0$\%$}	& 207 {\tiny$\pm$1.34$\%$}  & 194 {\tiny$\pm$1.88$\%$}  \\
    4500 	& 141 {\tiny$\pm$0.9$\%$}	& 1.11e-3		& 2.62e4 {\tiny$\pm$1.0$\%$}	& 186 {\tiny$\pm$1.34$\%$}  & 174 {\tiny$\pm$1.88$\%$}  & 1.11e-3	& 2.62e4 {\tiny$\pm$1.0$\%$}	& 186 {\tiny$\pm$1.34$\%$}  & 174 {\tiny$\pm$1.88$\%$}  & 1.01e-3	& 2.87e4 {\tiny$\pm$1.0$\%$}	& 204 {\tiny$\pm$1.34$\%$}  & 191  {\tiny$\pm$1.88$\%$}  \\
    5000 	& 156 {\tiny$\pm$0.9$\%$}	& 9.51e-4		& 3.05e4 {\tiny$\pm$1.0$\%$}	& 195 {\tiny$\pm$1.34$\%$}  & 182 {\tiny$\pm$1.88$\%$}  & 1.01e-3	& 2.86e4 {\tiny$\pm$1.0$\%$}	& 183 {\tiny$\pm$1.34$\%$}  & 171 {\tiny$\pm$1.88$\%$}  & 9.00e-4	& 3.23e4 {\tiny$\pm$1.0$\%$}	& 206 {\tiny$\pm$1.34$\%$}  & 193 {\tiny$\pm$1.88$\%$}  \\
    5500 	& 172 {\tiny$\pm$0.9$\%$}	& 9.23e-4		& 3.14e4 {\tiny$\pm$1.0$\%$}	& 183 {\tiny$\pm$1.34$\%$}  & 171 {\tiny$\pm$1.88$\%$}  & 		& 					& 				   & 	& 8.10e-4	& 3.58e4 {\tiny$\pm$1.0$\%$}	& 208 {\tiny$\pm$1.34$\%$}  & 195 {\tiny$\pm$1.88$\%$}  \\
    6000 	& 188 {\tiny$\pm$0.9$\%$}	& 8.20e-4		& 3.54e4 {\tiny$\pm$1.0$\%$}	& 189 {\tiny$\pm$1.34$\%$}  & 176 {\tiny$\pm$1.88$\%$}  & 8.11e-4	& 3.58e4 {\tiny$\pm$1.0$\%$}	& 191 {\tiny$\pm$1.34$\%$}  & 178 {\tiny$\pm$1.88$\%$}  & 7.37e-4	& 3.94e4 {\tiny$\pm$1.0$\%$}	& 210 {\tiny$\pm$1.34$\%$}  & 196 {\tiny$\pm$1.88$\%$}  \\
    6500 	& 203 {\tiny$\pm$0.9$\%$}	&			& 					& 				   &				      & 7.41e-4	& 3.92e4 {\tiny$\pm$1.0$\%$}	& 193 {\tiny$\pm$1.34$\%$}  & 180 {\tiny$\pm$1.88$\%$}  & 6.74e-4	& 4.31e4 {\tiny$\pm$1.0$\%$}	& 212 {\tiny$\pm$1.34$\%$}  & 198 {\tiny$\pm$1.88$\%$}  \\
    7000 	& 219 {\tiny$\pm$0.9$\%$}	& 			& 					& 				   & 				      & 6.92e-4	& 4.19e4 {\tiny$\pm$1.0$\%$}	& 192 {\tiny$\pm$1.34$\%$}  & 179 {\tiny$\pm$1.88$\%$}  & 6.22e-4	& 4.66e4 {\tiny$\pm$1.0$\%$}	& 213 {\tiny$\pm$1.34$\%$}  & 199 {\tiny$\pm$1.88$\%$}  \\
    					\hline
    					\hline
    \end{tabularx}}
    \end{sidewaystable}

    \end{document}